\documentclass[aps,pre,superscriptaddress,footinbib,showpacs,twocolumn]{revtex4-1}

\usepackage{amssymb,latexsym,mathrsfs}
\usepackage{amsfonts}
\usepackage{graphicx}
\usepackage{lmodern}
\usepackage{fix-cm}
\usepackage{xfrac}
\usepackage{color}    
\usepackage{lipsum}
\usepackage{textcomp}
\usepackage{amsmath}
\usepackage{mathtools}
\usepackage{subfigure}
\usepackage{natbib}
\usepackage{float}
\usepackage{datetime}
\usepackage{hyperref}
\hypersetup{
    colorlinks=true,
    linkcolor=red,
    filecolor=magenta,      
    urlcolor=cyan,
}
\usepackage{tabularx,ragged2e,booktabs}

\newcommand{\be}{\begin{equation}}
\newcommand{\ee}{\end{equation}}
\newcommand{\bearr}{\begin{array}}
\newcommand{\enarr}{\end{array}}

\newcommand{\eps}{\varepsilon}

\def\bea{\begin{eqnarray}}
\def\eea{\end{eqnarray}}
\def\ba{\begin{array}}
\def\ea{\end{array}}

\definecolor{dgreen}{rgb}{0,0.7,0}

\begin{document}

\title{Counter-flow induced clustering: Exact results}

\author{Amit Kumar Chatterjee}\email{ak.chatterjee@yukawa.kyoto-u.ac.jp}
\affiliation{Yukawa Institute for Theoretical
Physics, Kyoto University, Kitashirakawa Oiwakecho, Sakyo-ku, Kyoto 606-8502, Japan}
\author{Hisao Hayakawa}\email{hisao@yukawa.kyoto-u.ac.jp}
\affiliation{Yukawa Institute for Theoretical
Physics, Kyoto University, Kitashirakawa Oiwakecho, Sakyo-ku, Kyoto 606-8502, Japan}
\affiliation{Center for Gravitational Physics and Quantum Information,
Yukawa Institute for Theoretical Physics, Kyoto University, Kitashirakawa Oiwakecho, Sakyo-ku,
Kyoto 606-8502, Japan}

\begin{abstract}
We analyze the cluster formation in a non-ergodic stochastic system as a result of counter-flow,  with the aid of an exactly solvable model. 
To illustrate the clustering, a two species asymmetric simple exclusion process with impurities on a periodic lattice is
considered, where the impurity can activate flips between the two non-conserved species. Exact analytical results, supported by Monte Carlo simulations, show two distinct phases, {\it free flowing phase} and {\it clustering phase}. The clustering phase is characterized by constant density and vanishing current of the non-conserved species, whereas the free flowing phase is identified with non-monotonic density and non-monotonic finite current of the same.
The $n$-point spatial correlation between $n$ consecutive vacancies grows with increasing $n$ in the clustering phase, indicating the formation of two macroscopic clusters in this phase, one of the vacancies and the other consisting of all the particles.
We define a {\it rearrangement parameter} that permutes the ordering of particles in the initial configuration, keeping all the input parameters fixed. This rearrangement parameter reveals the significant effect of non-ergodicity on the onset of clustering. 
For a special choice of the microscopic dynamics, we connect the present model to a system of run-and-tumble particles used to model active matter, where the two species having opposite net bias manifest the two possible run directions of the run-and-tumble particles, and the impurities act as tumbling reagents that enable the tumbling process.  
\end{abstract}

\maketitle

\section{Introduction}
\label{intro}
Clustering of non-attractive elements is an intriguing  phenomenon occurring in diverse areas of science and society. It appears in various fields such as granular materials \cite{Liu_1998,Biroli_2007,Behringer_2019} , vehicular and pedestrian traffic flows \cite{Nagel_1992,Helbing_2001,Nagatani_2002,Schadschneider_2011}, active matter \cite{Mandal_2020,Yang_2022}, biology \cite{Sadati_2013,Oswald_2017} etc. At the heart of the clustering phenomena lies a transition between free flowing phase and jammed phase where motion becomes highly restrictive, as some suitable system parameter is tuned. The tuning parameter for granular material is packing fraction,  it is car or pedestrian density for traffic flow and self-propulsion force in case of dense active matter. In fact, a jamming phase diagram has been proposed keeping in mind the generality of such transitions \cite{Liu_1998} and there are experiments conducted on colloidal particles supporting the concept of this generic phase diagram such as 
Ref.~\cite{Trappe_2001}. The control of clustering is important in daily life.  Indeed, understanding the formation of jamming and finding ways to transit to unjammed state, has immense importance in traffic science. The other applications include occurrence of cellular jamming transitions in cancer \cite{Sadati_2013,Oswald_2017}. Analysis of clustering, as a phase transition in interacting many body systems, constitutes an interesting topic from the physics point of view. Thus, it seems natural to investigate the clustering phenomena through the lens of statistical mechanics.

One possible mechanism of clustering is the presence of counter-flow in a system. Counter-flows arise naturally in real life situations, e.g. pedestrian traffic flow \cite{Schadschneider_2011} in narrow lanes, busy pedestrian zones, crossings between two footpaths etc. In these situations, we often observe pedestrians moving in opposite  directions, automatically create counter-flow situations. Counter-flows in pedestrian dynamics can result in interesting phenomena like traffic jam at high densities \cite{Muramatsu_1999,Tajima_2002}, self-organized lane formations \cite{Helbing_1995,Hoogendoorn_2005}, oscillatory changes in the dominant direction of motion at bottlenecks \cite{Kretz_2006_1}; as well as unexpected events like panic and crowd disasters \cite{Klingsch_2010}. Interesting experimental results regarding counter-flows have revealed that the total flow in a counter-flow situation can be greater than the sum of the two comparable unidirectional flows \cite{Hoogendoorn_2005,Kretz_2006}. Numerical studies show that the existence of counter-flowing agents ignoring traffic rules can surprisingly lower the possibility of traffic jam \cite{Baek_2009}. Apart from the pedestrian dynamics, counter-flow occurs in nature and has been incorporated in industries. In fact, counter-current exchange of heat or chemicals between two oppositely flowing fluids (i.e. counter-flow situation), has been proved to be much more efficient than co-current exchange of the same between two parallel flowing fluids. Such principles are often used to devise heat exchangers \cite{Sridhar_2017} used in industries and are found in nature e.g. salt glands in sea and desert birds \cite{Schmidt_1959,Proctor_1993}, mammalian kidneys \cite{Gottschalk_1958} etc.

In the context of cluster formation resulting from counter flow, spontaneous symmetry breaking in a model with two oppositely moving particles with exchange interaction, has been discussed with periodic boundaries \cite{Lee_1997,Arndt_1998} and open boundaries \cite{Evans_1995}. Notably, later it has been proved that the spatial condensation of particles under periodic boundary condition \cite{Arndt_1998}, is not associated with a phase transition in the grand canonical ensemble \cite{Rajewsky_2000,Sasamoto_2001}.   
There exist several models in non-equilibrium statistical mechanics that exhibit phase transitions in one dimension \cite{Schmittmann_1994,Privman_1997,Marro_1999,Schutz_2001,Henkel_2008,Henkel_2010,Derrida_1993,Evans_1998,Lahiri_1997,Evans_2002,Mallick_1996,Derrida_1999}.
However, considerably few one dimensional non-equilibrium models with simplified local dynamics, are amenable to exact analytical calculations which provide much insights to the microscopic origins of the phase transitions \cite{Derrida_1993,Mallick_1996,Derrida_1999,Evans_2000,Mukamel_2000}. 
In context of cluster formations in exactly solvable disordered systems, Bose-Einstein condensates have been studied in multi-species asymmetric simple exclusion processes \cite{Evans_1996,Krug_1996},  utilizing matrix product ansatz \cite{Blythe_2007} and zero range process \cite{Spitzer_1970,Evans_2005}. Traffic jam in more realistic traffic models like bus route models, has been considered \cite{Loan_1998_1,Loan_1998_2} and its connection to Nagel-Schreckenberg model of traffic flow, has been explored in details \cite{Chowdhury_2000}. There are various other interesting studies of one dimensional traffic flows under periodic boundary conditions \cite{Nagel_1992,Helbing_2001,Nagatani_2002,Schadschneider_2011,Bando_1995,Komatsu_1995,Hayakawa_1998,Igarashi_1999}. Notably, once a formed cluster is stable and moves in one-direction, it is similar to the time-crystal \cite{Wilczek_2012,Shapere_2012,Watanabe_2015,Yao_2017,Else_2020}. To avoid such clustering, model with bypassing defects through long range hopping, has been analyzed \cite{Nossan_2011}. Notably, exact analytical results showing  clustering phenomenon has been considerably few. 

Recently, clustering of self-propelled objects like bacteria have gained much attention \cite{Berg_2004,Brumley_2019}. Such active matter can exhibit motility induced phase separation by  accumulating in regions with slow movement \cite{Tailleur_2008,Cates_2015,Fodor_2017}, or they can aggregate near chemical nutrients \cite{Brumley_2019}. The motion of self-propelled objects including bacteria, is often described by run-and-tumble particles (RTPs) \cite{Thompson_2011,Soto_2014,Slowman_2016,Malakar_2018,Mallmin_2019,Das_2020,Mukherjee_2022,Jose_2022}. Analytical solutions of run-and-tumble models are few, including exact solution for one and two RTPs \cite{Slowman_2016,Malakar_2018,Mallmin_2019,Das_2020,Mukherjee_2022}, mean field analysis of many interacting RTPs \cite{Dandekar_2020} and approximate solution for restricted RTPs \cite{Mukherjee_2022}. The tumbling dynamics corresponds to occasional change in direction of motion, which specifically in one dimension, would correspond to {\it flip} between right and left moving objects. Thus, it would be interesting to ask how to accommodate such flip dynamics of RTPs in simple lattice models that are exactly solvable,  contain many interacting RTPs and exhibit clustering.

In this work, we provide exact results showing cluster formation on a one dimensional periodic lattice, when counter-flow and flip dynamics are present in the system. As a model, we consider the two species asymmetric simple exclusion process with impurity activated flips ($2$-ASEP-IAF) \cite{Chatterjee_2022}. We show the existence of two different phases, namely the {\it free flowing phase} and the {\it clustering phase}, and characterize them using observables like average density, drift current and spatial correlations, obtained analytically and supported by Monte Carlo simulations. Interestingly, in the counter-flow situation, with a special choice of the microscopic hop rates, the $2$-ASEP-IAF can be interpreted as a system of many RTPs in presence of tumbling reagents. Since the RTPs have been extremely useful to model active matter on lattices, we would discuss separately the clustering in this special case in some details. Also, the $2$-ASEP-IAF has plausible mappings to two-lane ASEP with bridges \cite{Chatterjee_2022} and enzymatic chemical reactions \cite{Chatterjee_2022}, explained briefly in the next section. 

In Section~\ref{model} we define the model and briefly describe its connection to the two-lane model, enzymatic chemical reactions and RTPs. Section~\ref{ss} describes the steps to obtain the exact matrix product steady state of the model with infinite dimensional matrices. The model being non-ergodic, we state the choice of initial configuration considered here in Section~\ref{init} and calculate the partition function showcasing the important steps. In Section~\ref{two} we discuss in details the analytical results for density, current and spatial correlations which establish the existence of two different phases. The role of initial configurations on the onset of the clustering, is investigated in Section~\ref{non-ergodic}. The special constraint on the microscopic rates, for which the $2$-ASEP-IAF can be mapped to a model of run-and-tumble particles, is analyzed in details in Section~\ref{rtp}. We summarize our main results and future directions in Section~\ref{summary}. Appendix~\ref{app:mpa} provides expressions and brief derivations of some observables. The explicit form of the density-fugacity relation and a special case that leads to a closed form solution of the fugacity, are presented in Appendix~\ref{app:rhoz}. In Appendix~\ref{app:init} we clarify on the variations of the initial configuration considered in the main text, that can give rise to similar clustering. We show the convergence of our results with system size in Appendix~\ref{app:L}. The comparison of our model to the Arndt-Heinzel-Rittenberg model of counter-flow, is briefly discussed in Appendix~\ref{app:ahr}.
\section{Model}
\label{model}
We consider two different species ($1$ and $2$) along with impurities ($+$) and vacancies ($0$) on a one dimensional periodic lattice with $L$ sites, $i=1,2,\dots,L$. Species $1$ and species $2$ can hop to right with rates $p_1$ and $p_2$ respectively, or to left with rates $q_1$ and $q_2$ respectively, if the target site is vacant. The motion of the impurity is restricted only towards right, with rate $\epsilon$. The unidirectional motion of the impurity is a characteristic that differentiates it from the two species. Further, this assumption turns out to be crucial to obtain the exact steady state probability distribution. Alongside the hopping dynamics, species $1$ and $2$ can transform into each other with rates $w_{12}$ and $w_{21}$, such flipping being activated only in presence of impurity.  Following the nomenclature of Ref.~\cite{Chatterjee_2022}, we refer this model as $2$-ASEP-IAF, where ASEP represents asymmetric simple exclusion process and IAF stands for impurity activated flips.   The microscopic dynamics of $2$-ASEP-IAF is represented as
\begin{eqnarray}
 10 \,\,\xrightleftharpoons[q_{1}]{p_{1}}\,\,  01,  \hspace*{0.8 cm} &&
 20 \,\,\xrightleftharpoons[q_{2}]{p_{2}}\,\,  02,  \nonumber \\
 +0 \,\,\stackrel{\epsilon}{\longrightarrow}\,\,  0+,  \hspace*{0.8 cm}  &&
1+\,\, \xrightleftharpoons[w_{21}]{w_{12}}\,\, 2+ .
 \label{eq:dynamics}
\end{eqnarray}
The $2$-ASEP-IAF can be restated as a four-state model where each lattice site can be in either of the four possible states $1$ or $2$ or $+$ or $0$.
The input parameter space for the model contains several parameters, more precisely given by $(p_1, p_2, q_1, q_2, \epsilon, w_{12},w_{21},\rho_0,\rho_+)$. The parameters $\rho_0=N_0/L$ and $\rho_+=N_+/L$ are the conserved densities of the vacancies and the impurities,  respectively, with $N_0$ and $N_+$ being the number of vacancies and impurities in the system. In this paper, we aim to investigate the effect of counter-flow on the system, and to do so,
we pick out $q_1$ to be the tuning parameter, keeping all other rates fixed. We fix $p_2>q_2$ (apart from the discussion in Section~\ref{rtp}) and $\epsilon>0$. Subsequently, if $q_1<p_1$, all particles have net bias along the same direction, keeping the system in {\it natural flow} situation. On the other hand, when $q_1>p_1$, species $1$ has net bias in the direction opposite to the net bias of species $2$ and impurities, thereby species $1$ opposes the motion of other components and creates a {\it counter-flow} situation. Since the tuning parameter $q_1$ can control the flow situation in the system, we denote it as the {\it counter-flow parameter}.

We would like to mention three interesting plausible mappings of the present model [Eq.~(\ref{eq:dynamics})] to other systems of interest. These three systems, as briefly explained below, are (i) run-and-tumble particles, (ii) enzymatic chemical reaction and (iii) two-lane ASEP. 

(i) To study active matter on lattices, run-and-tumble particles (RTPs) constitute useful models \cite{Soto_2014}, notably with considerably few analytical results  \cite{Slowman_2016,Malakar_2018,Mallmin_2019,Das_2020,Mukherjee_2022,Jose_2022}. The RTPs keep moving along a particular direction until they tumble i.e. change the direction of motion. The two different species having net bias towards opposite directions in $2$-ASEP-IAF, can be considered as two possible orientations of the RTP, only if the counter-flow situation is considered. Basically the RTP is a two-state particle where its two possible states or run directions are imitated by species $1$ and $2$ of the $2$-ASEP-IAF. A particularly relevant case is when $p_2-q_2=q_1-p_1$ i.e. the RTP moves with same speed in both directions with intermediate tumbling, in accordance with the recently proposed restricted tumbling model \cite{Mukherjee_2022} and the continuum active random walk model \cite{Jose_2022}. Notably, in our mapping, the impurities act as the mediators for the tumbling process. This special case which maps the $2$-ASEP-IAF to RTPs, would be discussed in more details in Section~\ref{rtp}. 

(ii) The $2$-ASEP-IAF can be connected to an enzymatic chemical reaction in a narrow channel of diffusing chemical reagents. In this case, the impurity represent the enzymes which initiate the reaction between substrate and products, that amounts to the flip dynamics in the present model \cite{Chatterjee_2022}. 

(iii) The $2$-ASEP-IAF can be mapped to a two-lane ASEP, a simple model for two-lane traffic flow, where the species $1$ and $2$ play the roles of particles hopping with different rates in the two different lanes and the impurities in $2$-ASEP-IAF mimic the bridges connecting the two lanes in the two-lane ASEP \cite{Chatterjee_2022}. Thus the flip dynamics between two species activated by impurities represent the lane change dynamics of the particles in the two-lane ASEP model.

Due to such connections of the $2$-ASEP-IAF to several other important models, we would like to study analytically the possibility of cluster formation in the model in presence of counter-flow, which can possibly provide information about clustering in the connected models.

\section{Steady state}
\label{ss}
The probability $P(\left\lbrace s_i \right\rbrace)$ of any configuration $\left\lbrace s_i \right\rbrace$ ($s_i=1$ or $2$ or $0$ or $+$, denoting the constituent at site $i$), in the non-equilibrium steady state corresponding to the microscopic dynamics Eq.~(\ref{eq:dynamics}), is obtained in the following matrix product form
\begin{eqnarray}
P(\left\lbrace s_i \right\rbrace) &\propto& \mathrm{Tr}\left[\prod_{i=1}^{L}X_i\right], \cr
\mathrm{where}\hspace*{0.4 cm} X_i&=&D_1 \delta_{s_i,1}+D_2 \delta_{s_i,2}+A \delta_{s_i,+}+E \delta_{s_i,0}.
\label{eq:ss} 
\end{eqnarray}
In Eq.~(\ref{eq:ss}), the matrix $X_i$ represents the component $s_i$ at site $i$ and $\delta_(.,.)$ is  Kronecker delta symbol. Specifically, the matrices $D_1,D_2,E,A$ correspond to species $1$, species $2$, vacancy and impurity, respectively. The configurations of the system evolve according to the Master equation 
\be
\frac{d|P(t)\rangle}{dt}=M|P(t)\rangle,
\label{eq:sme}
\ee
where the matrix $M$ is the rate matrix containing transition rates between configurations and 
$|P(t)\rangle$ is the column vector whose elements are time dependent probabilities $ P\left(\left\lbrace s_i \right\rbrace,t\right)$ for all possible configurations $\left\lbrace s_i \right\rbrace$. In the steady state, the probabilities $P\left(\left\lbrace s_i \right\rbrace,t\right)$ converge to the time independent values $P\left(\left\lbrace s_i \right\rbrace\right)$ mentioned in Eq.~(\ref{eq:ss}). For $2$-ASEP-IAF with two-site local dynamics [Eq.~(\ref{eq:dynamics})], the rate matrix can be decomposed as 
\be
M=\sum_{i=1}^{L} \left(I\otimes\dots I\otimes\mathcal{M}_{i,i+1}\otimes I\dots \otimes I\right),
\label{eq:mdecompose}
\ee
where $\mathcal{M}_{i,i+1}$ is a $16\times16$ matrix acting on the pair of sites $(i,i+1)$ and $I$ is $4\times4$ identity matrix placed at every site except the pair $(i,i+1)$. In steady state $M|P\rangle=0$. The steady state can be achieved using the following two-site flux-cancellation condition
\bea
\mathcal{M}_{i,i+1} \mathbf{X}_i \otimes \mathbf{X}_{i+1} &=& \tilde{\mathbf{X}}_i \otimes \mathbf{X}_{i+1} - \mathbf{X}_i \otimes \tilde{\mathbf{X}}_{i+1},
\label{eq:flux_cancel}
\eea
where 
\bea
\mathbf{X} &=& \left(E,A,D_1,D_2\right)^T, \cr
\tilde{\mathbf{X}} &=& \left(\tilde{E},\tilde{A},\tilde{D}_1,\tilde{D}_2\right)^T, \label{eq:flux_cancel2}
\eea
where $(.)^T$ denotes the transpose of the row vector $(.)$ and $\tilde{E},\tilde{A},\tilde{D}_{1}.\tilde{D}_2$ are auxiliary matrices that are introduced to satisfy the steady state equation and these have to be found out consistently along with the matrix representations for $E,A,D_{1},D_2$. We find that suitable choices for the auxiliary matrices in this case, are 
\bea
\tilde{E}=1,\,\, \tilde{A}=0,\,\, \tilde{D}_1=0, \,\, \tilde{D}_2=0. 
\label{eq:auxiliaries}
\eea
Using the above choices of the auxiliary matrices in Eq.~(\ref{eq:flux_cancel}), we arrive at the following matrix algebra
\bea
p_1 D_1 E-q_1 E D_1 &=& D_1, \cr
p_2 D_2 E-q_2 E D_2 &=& D_2, \cr
\epsilon A E &=& A, \cr
w_{12} D_1 A&=&w_{21} D_2 A. 
\label{eq:asym-algebra}
\eea
We find that Eq.~(\ref{eq:asym-algebra}) is satisfied by the infinite dimensional matrix representations given below,
\begin{eqnarray}
E&=&\left(\begin{array}{cccccc}
         0 & 0 & 0 & 0 & . &. \\
         1 & 0 & 0 & 0 &. &. \\
         0 & 1 & 0 & 0 &. &. \\
         0 & 0 & 1 & 0 &. &. \\
         0 & 0 & 0 & 1 &\, &\, \\
         . & . &\, & \, &. & \, \\
         . & . &\, & \, &\, & .  \\
        \end{array}
\right), \hspace*{0.35 cm}
A=\left(\begin{array}{cccccc}
         1 & \frac{1}{\epsilon} & \frac{1}{\epsilon^2} & \frac{1}{\epsilon^3} & . &. \\
         0 & 0 & 0 & 0 &. &. \\
         0 & 0 & 0 & 0 &. &. \\
         . & . &. & . &. & . \\
         . & . &. & . &. & .  \\
        \end{array}
\right) \cr
D_I&=&\left(\begin{array}{cccccc}
         d_{I}^{1,1} & d_{I}^{1,2} & d_{I}^{1,3} & d_{I}^{1,4} & . &. \\
         0 & d_{I}^{2,2} & d_{I}^{2,3} & d_{I}^{2,4} &. &. \\
         0 & 0 & d_{I}^{3,3} & d_{I}^{3,4} &. &. \\
         0 & 0 & 0 & d_{I}^{4,4} &. &. \\
         . & . & \, & \, &. &\, \\
         . & . &\, & \, &\, & . \\
        \end{array}
\right), \hspace*{ 0.3 cm} I=1,2\cr
&& d_{I}^{m,m+r}=\frac{(m)_{r}}{r!\,p_I^r} \left(\frac{q_I}{p_I}\right)^{m-1} d_{I}^{1,1}, \hspace*{0.8 cm}\forall r\geqslant0\cr
&& d_1^{1,1} = w_{21}, \hspace*{0.5 cm}d_2^{1,1} = w_{12}.
\label{eq:matrices}
\end{eqnarray}
The notation $(m)_r$ corresponds to Pochhammer symbol for rising factorials, $(m)_r:=m(m+1)(m+2)\dots(m+r-1)$. In Eq.~(\ref{eq:matrices}), the matrices $D_1$ and $D_2$, representing the two non-conserved species, are upper triangular matrices, whose elements involve their corresponding hop rates and the flip rate from the other species (e.g. $D_1$ involves $p_1,q_1,w_{21}$). The impurity is represented by the matrix $A$ which has nonzero elements in the first row only, the elements being functions of the impurity hop rate $\epsilon$. The matrix $E$, characterizing vacancy, is a lower shift matrix.  

We should mention that the matrix representations for any number of non-conserved species $\mu$ ($\mu=2$ for the present discussion) i.e. $\mu$-ASEP-IAF, has been obtained recently in Ref.~\cite{Chatterjee_2022}.  However, we would like to emphasize that the exact analytical calculations of observables and their properties strongly depend on the choice of initial configurations, owing to the non-ergodicity resulting from the microscopic dynamics Eq.~(\ref{eq:dynamics}). Particularly, the initial configuration considered in Ref.~\cite{Chatterjee_2022} is only a special case of the one that would be discussed in the next section (Section~\ref{init}). In fact, interestingly, as we would show later, the onset of clustering and its demarcation from the free flowing phase, crucially depend on the choice of initial configuration (Section~\ref{non-ergodic}). 

\section{Initial configuration and partition function}
\label{init}
In spite of the presence of flip-dynamics, the microscopic dynamics in Eq.~(\ref{eq:dynamics}) preserves certain orderings of the different species and impurities from initial configuration. It implies that the system is non-ergodic and,  can access only a subspace of the whole configuration space, starting from a particular initial configuration. To demonstrate the cluster formation in $2$-ASEP-IAF, we consider the following initial configuration (represented by corresponding matrices),
\begin{equation}
D_2 A\dots D_2 A\,\,D_1 A\dots D_1 A\,\,D_1\dots D_1\,\,E\dots E.
\label{eq:typeIII}
\end{equation}
The dots in $\mathcal{Y}\dots \mathcal{Y}$ ($\mathcal{Y}=D_2A, D_1A, D_1$ and $E$) represent an uninterrupted sequence of the unit $\mathcal{Y}$. Note that, there are two types of species $1$ particles in Eq.~(\ref{eq:typeIII}). One type is those which can flip, belongs to the sequence $\mathcal{Y}=D_1A$, while the others are non-flipping species $1$ particles belonging to the sequence $\mathcal{Y}=D_1$. The density of such non-flipping species $1$ particles is denoted by 
\be
\bar{\rho}:=\frac{\bar{N}}{L}, 
\label{eq:rb}
\ee
where the number of particles of species $1$ that cannot flip to species $2$ is given by $\bar{N}$. We do not require to introduce such symbol for species $2$ since the number of non-flipping species $2$ particles is zero in Eq.~(\ref{eq:typeIII}). A careful look at the initial configuration [Eq.~(\ref{eq:typeIII})] and the dynamics [Eq.~(\ref{eq:dynamics})] reveals that each configuration in the system can be identified as a sequence of intervals between impurities. Each interval contains one species $1$ or species $2$ particle and vacancies, except one interval that contains additional $\bar{N}$ non-flipping species $1$ particles apart from the one particle that can flip. The species $1$ and species $2$ particles can hop to right or left within this interval and can flip into each other at the two boundaries (i.e. impurities) of the interval. Each interval can increase or decrease in size by the incoming or outgoing vacancies at the right and left interval boundaries, respectively. This effectively gives rise to a cyclic motion of the intervals towards right.

For convenience, we calculate the partition function corresponding to the initial configuration Eq.~(\ref{eq:typeIII})  in the grand canonical ensemble, by associating the fugacity $z_0$ with the vacancies ($0$). Consequently, the partition function would be
\begin{eqnarray}
&& Q= \sum_{m_1=0}^{\infty}\dots\sum_{m_{N_+}=0}^{\infty} \,\, \sum_{\bar{m}_1=0}^{\infty}\dots\sum_{\bar{m}_{N_+}=0}^{\infty} \,\,\sum_{n_1=0}^{\infty}\dots\sum_{n_{\bar{N}}=0}^{\infty} \mathrm{Tr}\cr  &&\left[\left(\prod_{i=1}^{N_+}(D_1+D_2)(z_0 E)^{m_i}A(z_0 E)^{\bar{m}_i}\right)\left(\prod_{j=1}^{\bar{N}}D_1 (z_0 E)^{n_j}\right)\right].\nonumber \\
\hspace*{-2.5 cm}\label{eq:pf1}
\end{eqnarray}
The emergence of the products inside the trace in Eq.~(\ref{eq:pf1}) can be understood from the interpretation of each configuration as a sequence of intervals between impurities discussed above. Since we have $N_+$ impurities in the periodic system, that leads to $N_+$ intervals each bounded by impurities at both ends. This corresponds to the product of $N_+$ intervals in the first term inside the trace in Eq.~(\ref{eq:pf1}). 
Each product in the first term inside the trace correspond to interval containing one particle that can flip, whereas the last product term denotes the sequence of $\bar{N}$ non-flipping species $1$ particles inside the one exceptional interval.
We would use the explicit matrix representations from Eq.~(\ref{eq:matrices}) to calculate the partition function in Eq.~(\ref{eq:pf1}).  It is suitable to write down the matrices in the concise form as
\begin{eqnarray}
E&=& \sum_{\gamma=1}^{\infty} |\gamma+1\rangle \langle \gamma| \Rightarrow E^n= \sum_{\gamma=1}^{\infty} |\gamma+n\rangle \langle \gamma|, \cr
D_{1,2}&=& \sum_{\alpha=1}^{\infty} \sum_{\beta=\alpha}^{\infty} (d_{1,2})_{\alpha,\beta} |\alpha\rangle \langle \beta|, \cr (d_{1,2})_{\alpha,\beta}&:=&\frac{(\beta-1)!}{(\alpha-1)!(\beta-\alpha)!}\, \frac{q_{1,2}^{\alpha-1}}{p_{1,2}^{\beta-1}}, \cr
A &=& \sum_{\delta=1}^{\infty} \frac{1}{\epsilon^{\delta-1}} |1\rangle \langle \delta|,
 \label{eq:pf2}
\end{eqnarray}
where $\langle k|=(0,0,\dots 1,\dots0)$ is a standard basis vector with only non-zero element $1$ at the $k$-th place and $|k\rangle=(0,0,\dots 1,\dots0)^{T}$ where the superscript $T$ denotes transpose of the vector under consideration. 
With the above expressions, we can simplify the matrix strings in Eq.~(\ref{eq:pf1}). For example, we obtain
\begin{eqnarray}
&& \prod_{j=1}^{\bar{N}}D_1 (z_0 E)^{n_j} = \sum_{n_1}\dots\sum_{n_{\bar{N}}}\,\,\sum_{\alpha_1}\sum_{\alpha_2}\dots \sum_{\alpha_{\bar{N}}}\,\sum_{\beta_{\bar{N}}} 
\,z_0^{n_1+\dots+n_{\bar{N}}}\,\cr &&(d_1)_{\alpha_1,\alpha_2+n_1}\dots (d_1)_{\alpha_{\bar{N}-1},\alpha_{\bar{N}}+n_{\bar{N}-1}} \, (d_1)_{\alpha_{\bar{N}},\beta_{\bar{N}}+n_{\bar{N}}} |\alpha_1\rangle \langle \beta_{\bar{N}}|.\nonumber \\
\label{eq:pf3}
\end{eqnarray}
We incorporate the explicit form of $(d_1)_{\alpha,\beta}$ from Eq.~(\ref{eq:pf2}) to evaluate the sums in Eq.~(\ref{eq:pf3}). However, it is instructive to perform the above calculation recursively, i.e. first for single $D_1$, then two $D_1$, followed by three $D_1$ and finally generalize the result for $\bar{N}$ $D_1$-s by noting the trend. On the other hand, after invoking Eq.~(\ref{eq:pf2}), the other string of matrices in Eq.~(\ref{eq:pf1}) reduces to
\begin{eqnarray}
&& \prod_{i=1}^{N_+}(D_1+D_2)(z_0 E)^{m_i}A(z_0 E)^{\bar{m}_i} =\cr  &&\left(\sum_{m_1}z_0^{m_1}\sum_{\alpha_1} \left((d_1)_{\alpha_1,1+m_1} + (d_2)_{\alpha_1,1+m_1}\right) |\alpha_1\rangle \right)\times \cr && \left(\sum_{m}\sum_{n}\frac{z_0^{m+n}}{\epsilon^{n-1}}\sum_{\alpha} \left( \frac{(d_1)_{\alpha,1+m}}{\epsilon^{\alpha}}+  \frac{(d_2)_{\alpha,1+m}}{\epsilon^{\alpha}}\right) \right)^{N_+-1}\times\cr &&\sum_{\bar{m}_{N_+}}z_0^{\bar{m}_{N_+}}\sum_{\delta_{N_+}} \frac{\langle \delta_{N_+} |}{\epsilon^{\delta_{N_+}-1+\bar{m}_{N_+}}}.
\label{eq:pf4}
\end{eqnarray}
Using Eqs.~(\ref{eq:pf3}) and (\ref{eq:pf4}) in Eq.~(\ref{eq:pf1}), along with the explicit forms of $(d_{1,2})_{\alpha,\beta}$ from Eq.~(\ref{eq:pf2}), we finally arrive at the following expression of the partition function 
\begin{eqnarray}
&& Q =\left(\frac{1}{1-\frac{z_0}{\epsilon}}\right)^{N_+}\left[\frac{w_{21}}{1-\frac{z_0}{p_1}-\frac{z_0}{\epsilon}\frac{q_1}{p_1}}+\frac{w_{12}}{1-\frac{z_0}{p_2}-\frac{z_0}{\epsilon}\frac{q_2}{p_2}}\right]^{N_+-1}\cr && \times \prod_{k=1}^{\bar{N}}\frac{w_{21}}{1-\frac{z_0}{p_1}S_{k-1}} \times \left[\frac{w_{21}}{1-\frac{z_0}{p_1}S_{\bar{N}}}+\frac{w_{12}}{1-\frac{z_0}{p_2}-\frac{q_2}{p_2}\frac{z_0}{p_1}S_{\bar{N}-1}}\right], \nonumber \\
 \label{eq:pf5}
\end{eqnarray}
where 
\begin{eqnarray}
S_k&:=&\sum_{j=0}^{k} \left(\frac{q_1}{p_1}\right)^{j}+\left(\frac{q_1}{p_1}\right)^{k}\frac{q_1}{\epsilon}\cr &=&\frac{p_1\left[\left(\frac{q_1}{p_1}\right)^{k+1}((k+1)(p_1-q_1)-\epsilon)+\epsilon\right]}{(p_1-q_1)\epsilon}.
 \label{eq:pf6}
\end{eqnarray}
The fugacity $z_0$ is computed from the density-fugacity relation 
\be
\rho_0=\frac{z_0}{L}\frac{d}{dz_0}\mathrm{ln}Q.
\label{eq:rhozapp}
\ee
Thus, we have obtained the analytical form of the partition function Eq.~(\ref{eq:pf6}) corresponding to the initial configuration Eq.~(\ref{eq:typeIII}).
\begin{figure}[t]
\centering
\subfigure[]{\includegraphics[scale=0.65]{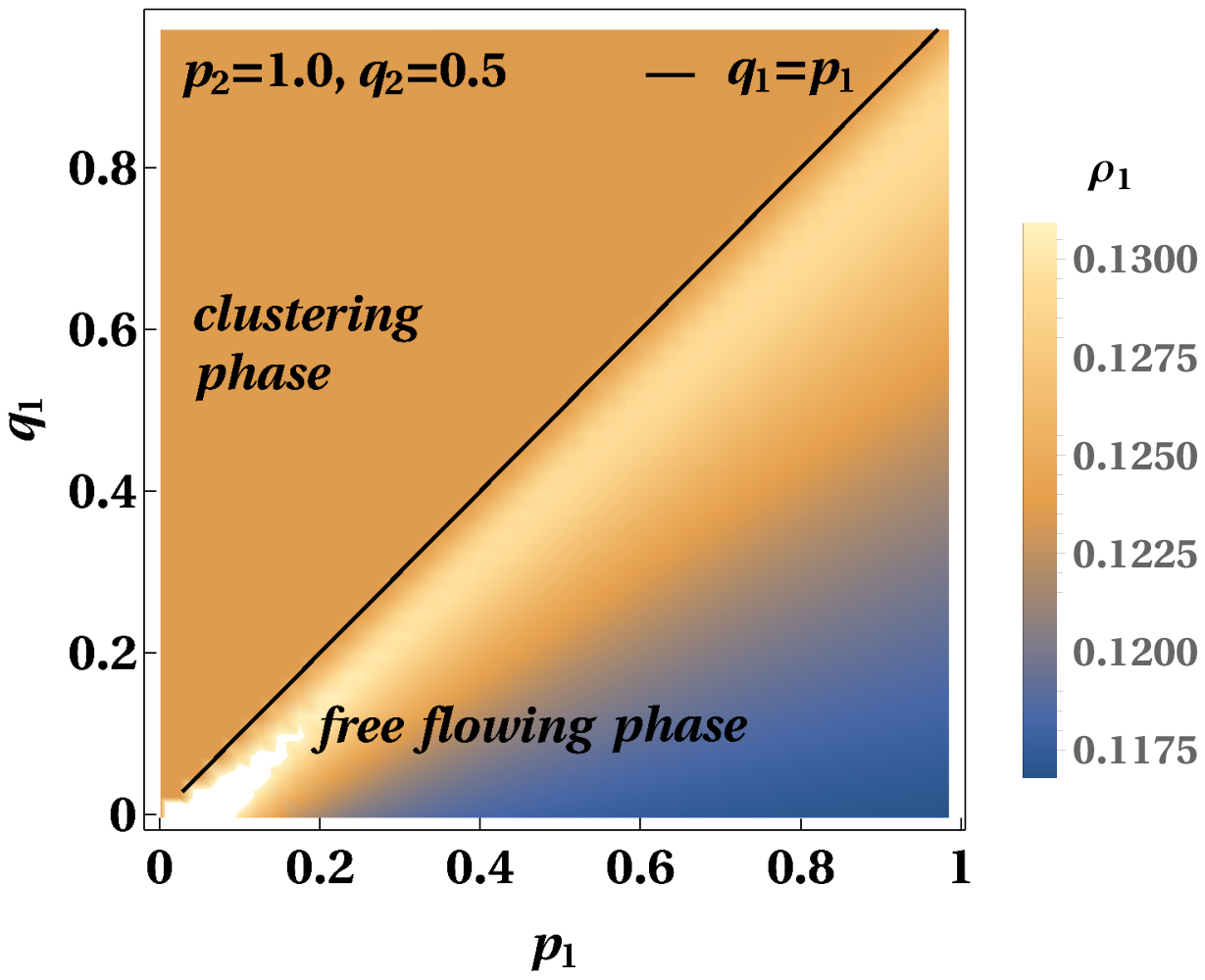}}\\
\subfigure[]{\includegraphics[scale=0.65]{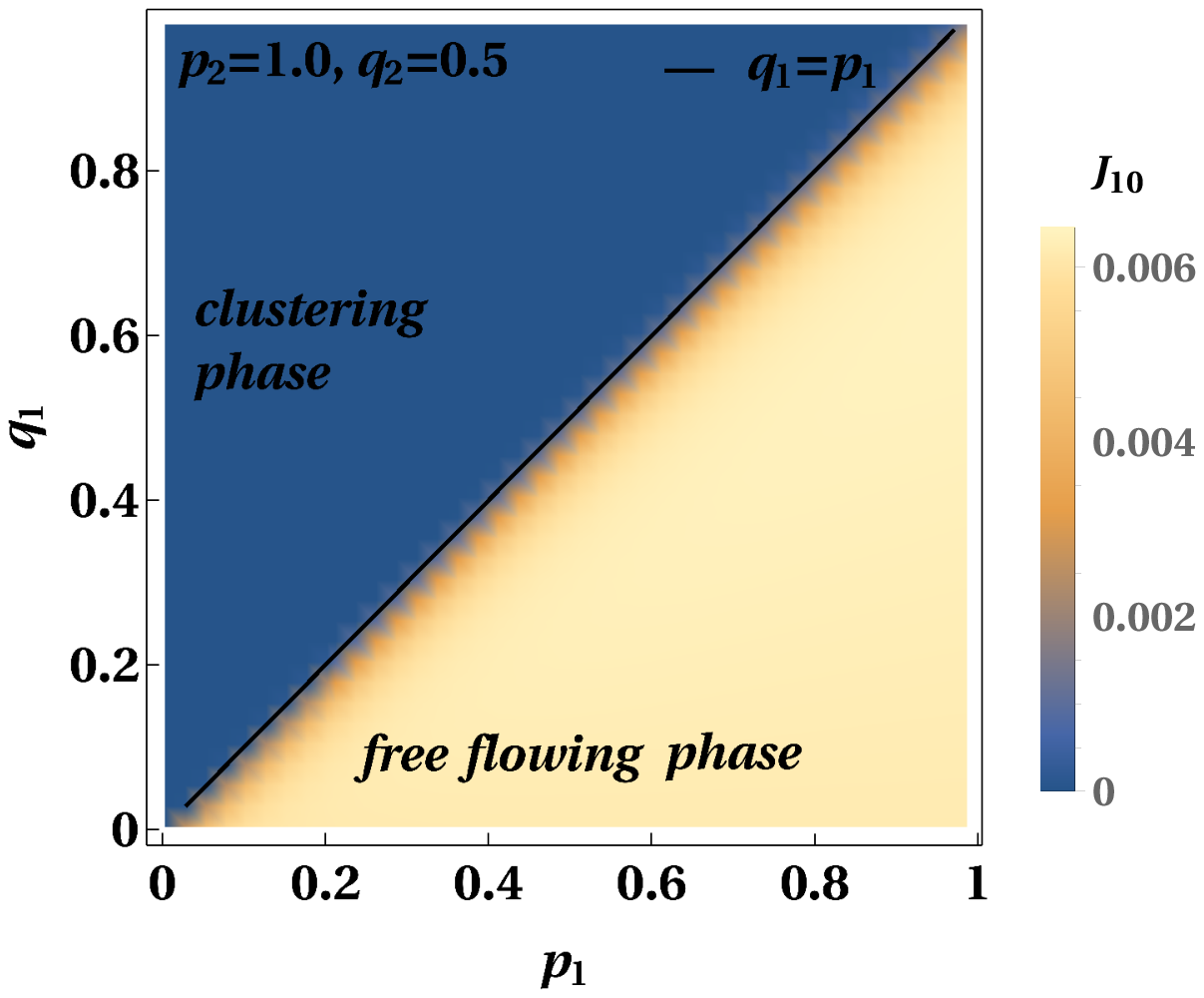}}\\
\caption{The figures (a) and (b) show heat-maps for density $\rho_1$ and current $J_{10}$ respectively, for species $1$, in the $p_1-q_1$ parameter plane. Both of them, particularly figure (b) clearly exhibits that $q_1=p_1$ separates two different phases. Figure (a) also implies the non-monotonicity of the density. The parameters used are $L=10^3, p_2=1.0, q_2=0.5, \epsilon=0.1, w_{12}=1.0, w_{21}=0.1, \rho_+=0.25, \rho_0=0.4$.}.
\label{fig:heatmap}
\end{figure}
\section{Two phases: analytical results}
\label{two}
In this section, we would discuss exact results for observables, showing the emergence of two different phases in the $2$-ASEP-IAF, with the variation of the parameter $q_1$. 
Starting from the  initial configuration Eq.~(\ref{eq:typeIII}), the analytical calculations are performed following the footsteps sketched in the previous section, where we use the explicit representations of the matrices from Eq.~(\ref{eq:matrices}).
Before entering into the detailed discussions of the observables, below we briefly summarize our main findings.\\
(i) The analytical results for one-point (average species densities), two-point (drift current) and $n$-point functions (correlation between consecutive vacancies) exhibit the existence of two-different phases: the {\it free flowing phase} and the {\it clustering phase}. In Figs.~\ref{fig:heatmap}(a) and (b), we present the heat maps for average species density and average drift current (for species $1$) respectively, in the parameter plane $p_1-q_1$. Fig.~\ref{fig:heatmap}(b) prominently distinguishes two phases separated by the line $q_1=p_1$. Fig.~\ref{fig:heatmap}(a) also shows two distinct phases demarcated by a region around the line $q_1=p_1$.\\ 
(ii) The average species density (of species $1$), when plotted against $q_1$, remarkably exhibits two different behaviors for $q_1<p_1$ and $q_1>p_1$. For $q_1<p_1$, defining the free flowing phase, the density is {\it non-monotonic}, whereas it remains {\it constant} for  $q_1>p_1$ that constitutes the clustering phase. Thus the non-conservation plays an important role in identifying the two different phases through a simple one point function. \\ 
(iii) The average drift current (of species $1$) is {\it non-zero} in the free-flowing phase, followed by a sharp descent toward zero near $q_1=p_1$, and remains vanishingly small in the clustering phase. \\
(iv) The $n$-point correlation function between $n$ consecutive vacancies increases with increasing $n$ in the counter-flow phase, directly pointing towards the formation of vacancy cluster, and consequently that of particle cluster. 
\subsection{Average species density}
\label{sub:rho}
\begin{figure}[t]
  \centering \includegraphics[width=8.6 cm]{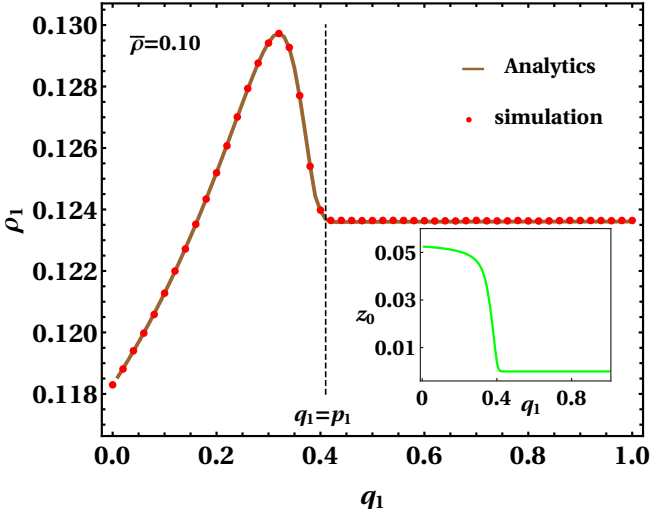}
  \caption{The figure shows the behavior of the average species density as a function of the counter-flow parameter, obtained from analytical result (brown solid line) and Monte Carlo simulation (red dots). Two different phases are apparent, separated by the transition point $q_1=p_1$. For $q_1<p_1$, the density exhibits non-monotonicity whereas for $q_1>p_1$ the density remains constant. The inset presents the variation of $z_0$ with $q_1$. The fugacity exhibits a sharp decrease towards zero in $q_1\lesssim p_1$ and remains close to zero for $q_1>p_1$.} The parameters used are $L=10^3, p_1=0.4, p_2=1.0, q_2=0.5, \epsilon=0.1, w_{12}=1.0, w_{21}=0.1, \rho_+=0.25, \rho_0=0.4$..
\label{fig:density}
\end{figure}

We first consider the one-point functions, i.e. the average species densities $\rho_1=\langle 1 \rangle$ and $\rho_2=\langle 2 \rangle$ of the non-conserved species $1$ and $2$, respectively, where $\langle . \rangle$ denotes ensemble average in the steady state. Since, the total density of the two species remain constant ($\rho_1+\rho_2=\bar{\rho}+\rho_+$), it would be sufficient to focus only on the density behavior of one of the species, say $\rho_1$.  The formal expression for $\langle 1 \rangle$ that has been used to calculate $\rho_1$, is presented in Eq.~(\ref{eq:ob1}) of Appendix~\ref{app:mpa}. With the aid of Eqs.~(\ref{eq:pf3}) and (\ref{eq:pf4}), the Eq.~(\ref{eq:ob1}) leads to the following exact formula
\bea
\rho_1&=&\bar{\rho}+\left(\rho_+-\frac{1}{L}\right)\frac{\frac{w_{21}}{1-\frac{z_0}{p_1}S_0}}{\left[\frac{w_{21}}{1-\frac{z_0}{p_1}S_0}+\frac{w_{12}}{1-\frac{z_0}{p_2}\left(1+\frac{q_2}{p_1} S_{-1}\right)}\right]}\cr && +\frac{1}{L} \frac{\frac{w_{21}}{1-\frac{z_0}{p_1}S_{\bar{N}}}}{\left[\frac{w_{21}}{1-\frac{z_0}{p_1}S_{\bar{N}}}+\frac{w_{12}}{1-\frac{z_0}{p_2}\left(1+\frac{q_2}{p_1} S_{\bar{N}-1}\right)}\right]},
\label{eq:rho}
\eea
where $S_k$ ($k=0$ or $-1$ or $N$ or $N-1$) is given by Eq.~(\ref{eq:pf6}). To express the density in Eq.~(\ref{eq:rho}) as a function of the input parameters only, it is required to solve for $z_0$, the density-fugacity relation Eq.~(\ref{eq:rhozapp}) which more explicitly takes the form Eq.~(\ref{eq:srhozexp}). However, the closed form solution of  Eq.~(\ref{eq:srhozexp}) is difficult to obtain. The solution of $z_0$ obtained from Mathematica, is used to replace it in Eq.~(\ref{eq:rho}). We present the corresponding analytical results (solid line) for $\rho_1$, agreeing with Monte Carlo simulation results (dots) as a function of $q_1$ in Fig.~\ref{fig:density}. 
It is fascinating that even the simplest one-point function, namely the average species density, clearly indicates the existence of two different phases. In one phase, occurring in the parameter region $q_1<p_1$, the density exhibits a {\it non-monotonic} behavior. Whereas it remains {\it constant} in the other phase, characterizing the parameter region  $q_1>p_1$. As one increases $q_1$ starting from zero, the hopping of species $1$ particles to left, becomes increasingly likely. This means lesser chances for species $1$ particles to have impurities as their right nearest neighbors. Consequently, the flipping of species $1$ to $2$ decreases with increasing $q_1$, and therefore $\rho_1$ increases. After reaching the maximum density, $\rho_1$ starts decreasing, that results in a steep fall near $q_1\lesssim p_1$. For $q_1>p_1$, the density remains constant indicating that the drift process no longer can affect $\rho_1$, meaning the species $1$ particles cannot access vacancies due to possible clustering. The point $q_1=p_1$ demarcating two different phases, is regarded as the transition point. Note that the Fig.~\ref{fig:density} is a cross section of Fig.~\ref{fig:heatmap}(a) at $p_1=0.4$, and further clarifies the behavior of $\rho_1$ observed in the heat-map in Fig.~\ref{fig:heatmap}(a).

To see how the two different behaviors of $\rho_1$, non-monotonic (in $q_1<p_1$) and constant (in $q_1>p_1$), are influenced by that of the fugacity $z_0$, we show the variation of $z_0$ with $q_1$ in the inset of  Fig.~\ref{fig:density}. The inset shows that $z_0$ has a sharp decrease near $q_1=p_1$ and it is almost zero in the regime $q_1>p_1$. Consequently, we can take the limit $z_0\rightarrow0$ in Eq.~(\ref{eq:rho}) which would give us the working formula for $\rho_1$ in the constant density phase. This leads us to the following formula
\bea
\rho_1=\bar{\rho}+ \frac{w_{21}}{w_{12}+w_{21}}\rho_+, \hspace*{0.4 cm}\mathrm{for}\,\,q_1>p_1.
\label{eq:rho-constant}
\eea
The Eq.~(\ref{eq:rho-constant}) clearly manifests that $\rho_1$ essentially becomes independent of the parameter $q_1$ in the regime $q_1>p_1$. The exact above expression for the average species density can be obtained in a situation where we have two species $1$ and $2$ along with impurities in a system, but no vacancies, meaning the only dynamics is $1+ \xrightleftharpoons[w_{21}]{w_{12}} 2+$ and there is no drift since there are no vacancies. Such analysis strongly points towards the existence of a single cluster consisting of all the particles accompanied by another cluster consisting of only vacancies in our system in the parameter region $q_1>p_1$. Consequently, this phase is referred as the {\it clustering phase}. On the other hand, for $q_1<p_1$, the drift process is significant with flows of particles and vacancies, thereafter named as the {\it free flowing phase}. In the free flowing phase, near the transition point i.e. $q_1\lesssim p_1$, we find from Eq.~(\ref{eq:rho}) that the fall of $\rho_1$ towards the constant value [Eq.~(\ref{eq:rho-constant})] takes the form below
\bea
\rho_1&\approx&\bar{\rho}+\frac{w_{21}}{w_{12}+w_{21}}\rho_+\cr &&
+\, z_0\, \frac{w_{21}w_{12}}{(w_{12}+w_{21})^2} \left(A_1+A_2\right) \hspace*{0.2 cm}\mathrm{for}\,\,q_1\lesssim p_1,
\label{eq:rho-nonmon}
\eea
where  $A_1=\left(\frac{1}{p_1}+\frac{1}{\eps}-\frac{1}{p_2}-\frac{q_2}{p_2 \eps}\right)\left(\rho_+-\frac{1}{L}\right)$ and $A_2=\left(\frac{1}{p_1}+\frac{1}{\eps}\right)\left(\bar{\rho}(1-\frac{q_2}{p_2})+\frac{1}{L}\right)-\frac{1}{p_2 L}$, we have neglected higher orders of $z_0$. It is evident from Eq.~(\ref{eq:rho-nonmon}) that the fall of $\rho_1$ in the regime $q_1\lesssim p_1$ is linear in $z_0$ and would depend on how $z_0$ falls to zero in $q_1\lesssim p_1$ as a function of $(p_1-q_1)$. 
\begin{figure}[t]
  \centering \includegraphics[width=8.6 cm]{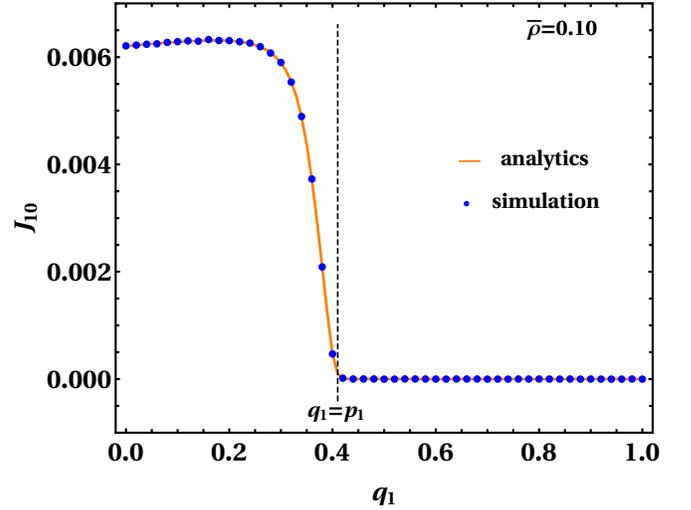}
  \caption{The figure demonstrates two different phases manifested by the drift current, obtained from analytical result (orange solid line) and Monte Carlo simulation (blue dots). In the parameter regime $q_1<p_1$, the current is finite whereas it falls abruptly to vanishingly small values at $q_1=p_1$ and remains so for $q_1>p_1$. The parameters used are $L=10^3, p_1=0.4, p_2=1.0, q_2=0.5, \epsilon=0.1, w_{12}=1.0, w_{21}=0.1, \rho_+=0.25, \rho_0=0.4$.}
\label{fig:current}
\end{figure}
\subsection{Average drift current}
\label{sub:j}
The drift currents for the non-conserved species and the impurity are given by
\bea
J_{10}=p_1\langle 10\rangle - q_1 \langle 01\rangle=z_0 \rho_1, \cr 
J_{20}=p_2\langle 10\rangle - q_2 \langle 01\rangle=z_0 \rho_2, \cr
J_{+0}=\epsilon\langle +0\rangle=z_0 \rho_+.
\label{eq:j}
\eea
The average steady state densities of the non-conserved species are provided in Eq.~(\ref{eq:rho}) whereas the impurity density $\rho_+$ is a conserved quantity. The fugacity $z_0$ obeys Eq.~(\ref{eq:rhozapp}), more precisely, is the solution of Eq.~(\ref{eq:srhozexp}). The analytical result for $J_{10}$ has been presented as a heat-map in Fig.~(\ref{fig:heatmap})(b) which prominently exhibits the existence of two difference phases, demarcated by the line $q_1=p_1$. In Fig.~\ref{fig:current}, we present a cross section of  Fig.~(\ref{fig:heatmap})(b) at $p_1=0.4$. 
We observe that the current, starting from non-zero value in the free flowing phase ($q_1<p_1$), shows a weak increase. This is due to the increasing density in this parameter regime (Fig.~\ref{fig:density}). However, the slope of the increase of $J_{10}$ (Fig.~\ref{fig:current}) is much smaller than that of $\rho_1$ (Fig.~\ref{fig:density}), because $J_{10}=z_0\rho_1$ and $z_0$ decreases with $q_1$ in the same parameter regime (inset of Fig.~\ref{fig:density}). After reaching the maximum, $J_{10}$ exhibits a sharp fall near the transition point $q_1=p_1$, and it remains almost zero (i.e. vanishingly small values) in the clustering phase ($q_1>p_1$). This observation is consistent with the emergence of the cluster formation for $q_1>p_1$ corresponding to the approximate situation where the vacancies merely play any role i.e. the drift process is almost absent. As mentioned in the previous section, this gives rise to two macroscopic clusters, one consisting of all particles and the other made of only vacancies. With the drift becoming negligible in the clustering phase, the density remains constant [Fig.~\ref{fig:density}, Eq.~(\ref{eq:rho-constant})] and the current stays vanishingly small [Fig.~\ref{fig:current}].
The species $2$ and impurity would have similar characteristics of current, therefore we restrict our investigation to $J_{10}$ only. 

From the analysis of the average species density and average current, we understand that the $2$-ASEP-IAF can exist in two distinct phases, the free flowing phase characterized by non-monotonic density and non-zero current, and the clustering phase identified by constant density and vanishing current. It is important to note that the transition occurs at $q_1=p_1$ meaning that sufficiently small counter-flow is enough to impose clustering in the system.
\subsection{Cluster formation}
In the previous sections \ref{sub:rho} and \ref{sub:j}, we have inferred the formation of clusters from the features of the density and the current. In this section, we discuss the physical origin of the clustering in $2$-ASEP-IAF and provide direct observational evidence of the cluster formation. To analyze the reason behind clustering, we remember that every configuration of the system can be interpreted as a sequence of intervals, as discussed in section \ref{init}. Each interval has two impurities at the interval boundaries and particles (species $1$ and species $2$) that drift within intervals and flip at the boundaries. The total number of particles within each interval is conserved. The intervals can drift towards right and increase or decrease in size as vacancies can enter into and exit from these intervals. The non-flipping species $1$ particles  inside the rightmost interval presented in Eq.~(\ref{eq:typeIII}), form a dense region against the impurity to their left. This is because the impurity can only move to right as opposed to the non-flipping species $1$ sequence that prefers to move left. For each interval, the vacancies can enter the interval {\it only through right boundary} whereas they can exit from the interval {\it only through left boundary}. Thus, whichever way the vacancies are distributed initially within the intervals, at long time they finally exit through the left boundaries of all the intervals and aggregate between the left boundary of leftmost interval and the accumulated species $1$ particles in the rightmost interval. Such clustering of vacancies also imply the formation of a cluster of all particles (species $1$ and species $2$ and impurities). Once the two clusters (particles and vacancies) are formed, the only way for re-entrance of the vacancies inside the particle cluster is by hopping through the accumulated non-flipping species $1$ particles. This requires right hopping of these species $1$ particles which is less probable as they are left biased. As the sequence of non-flipping species $1$ particles becomes larger, it takes longer time for a vacancy to travel thorough the whole particle cluster. It points towards a slow shift of the particle cluster to right and we expect the velocity of the shift to become smaller with increasing $\bar{\rho}$ [Eq.~(\ref{eq:rb})]. We denote the average velocity of the particle cluster by $v_{\mathrm{cl}}$. From the expressions for currents (as product of density and velocity) in Eq.~(\ref{eq:j}), the fugacity $z_0$ [Eq.~(\ref{eq:rhozapp})] is identified to $v_{\mathrm{cl}}$.
\begin{figure}[t]
  \centering \includegraphics[width=8.6 cm]{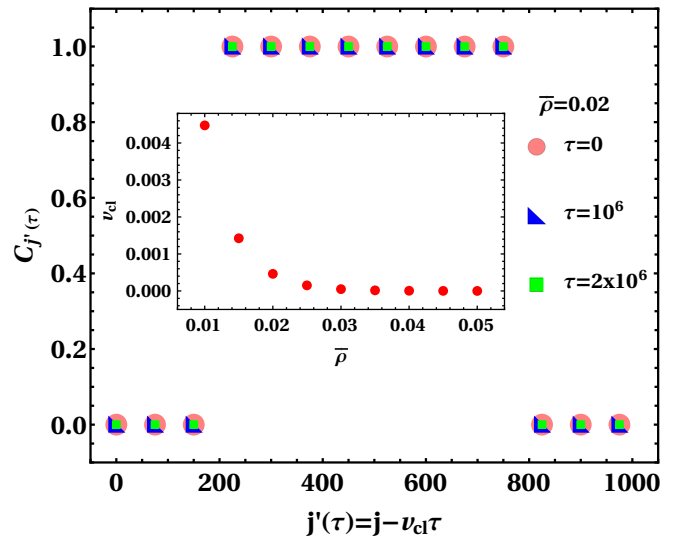}
  \caption{The figure shows the formation of two macroscopic clusters of particles and vacancies for $q_1>p_1$, obtained from Monte Carlo simulation. The clusters remain stationary in the shifted reference frame $j'=j-v_{\mathrm{cl}}\tau$, implying the drift of cluster with velocity $v_{\mathrm{cl}}$ in the original frame of reference. The cluster velocity $v_{\mathrm{cl}}$ decreases with increasing $\bar{\rho}$ [Eq.~(\ref{eq:rb})], shown in the inset. The parameters used are $L=10^3, p_1=0.4, q_1=0.5, p_2=1.0, q_2=0.5, \epsilon=0.1, w_{12}=1.0, w_{21}=0.1, \rho_+=0.29$ and $\rho_0=1-2\rho_+-\bar{\rho}$.}
\label{fig:cj}
\end{figure}

Next we show the clustering and the drift of the particle cluster explicitly. After the cluster is formed, we start our observation from any configuration and note the position of the particle cluster (determined by the impurity at the leftmost end of the particle cluster) at $\tau=0$. In Monte Carlo simulation, we measure the position of the cluster at different $\tau$. After time $\tau$, the cluster should shift by an amount of $v_{\mathrm{cl}}\tau$. It would be convenient to consider a frame of reference where the cluster remains stationary. Such stationary cluster frame of reference can be conceived by using the shifted lattice index $j'(\tau):=j-v_{\mathrm{cl}}\tau$. To characterize the position of particles and vacancies in any configuration, we consider the following observable in the shifted reference frame:
\begin{eqnarray}
C_{j'(\tau)}&=&1 \hspace*{0.5 cm} \mathrm{if}\,\, s_j'=1\,\, \mathrm{or}\,\, s_j'=2\,\,  \mathrm{or}\,\, s_j'=+ \cr
&=&0 \hspace*{0.5 cm} \mathrm{if}\,\, s_j'=0.
\label{eq:cj}
\end{eqnarray}
From the Monte Carlo simulation results in Fig.~\ref{fig:cj}, we observe that two macroscopic clusters are formed in the system at any $\tau$, one consisting of all the particles and the other made up of all the vacancies. Further, the cluster is observed to form at the {\it same} position in the shifted frame of reference, for different $\tau$. This implies that the cluster indeed moves with velocity $v_{\mathrm{cl}}$ in the original frame of reference. The slow shift of the cluster is similar to time crystals~\cite{Wilczek_2012,Shapere_2012,Watanabe_2015,Yao_2017,Else_2020}. 
Note that the slow shift in our model is caused by the unidirectional motion of impurities, and thus the emergence of this slow shift is not contradictory with the no-go theorem of time-crystals~\cite{Watanabe_2015}. The inset of Fig.~\ref{fig:cj} shows the variation of $v_{\mathrm{cl}}$ as a function of the density $\bar{\rho}$ [Eq.~(\ref{eq:rb})] of the non-flipping species $1$ particles. We observe that, with increasing $\bar{\rho}$, the cumulative effect of non-flipping species $1$ particles (prone towards left hopping) increases and thereby the cluster velocity (towards right) decreases.
\subsection{Spatial correlation}
An illuminating way to analytically show the formation of clusters, is to calculate the $n$-point correlation between consecutive vacancies (it is more helpful than calculating correlations between particles, because we have mixture of different species and impurities inside the particle cluster). The formal expression for the expression for $n$-point correlation function between $n$ consecutive vacancies, is the following
\be
C^{[n]}_0=\langle 00\dots0\rangle-\langle0\rangle^n.
\label{eq:n-point}
\ee
The superscript $[n]$ in Eq.~(\ref{eq:n-point}) represents the length of the uninterrupted sequence of consecutive vacancies (denoted by $0$ in the subscript).
In fact, the above correlation can be obtained recursively starting from the two-point correlation $C^{[2]}_0=\langle 00 \rangle-\rho_0^2$. We obtain the following expression for the two point correlation (see Appendix~\ref{app:mpa} for more details),
\bea
&& \hspace*{-0.2 cm} C^{[2]}_0=\rho_0 -\rho_0^2-\left(\rho_+-\frac{1}{L}\right)\frac{z_0}{\epsilon}-\frac{1}{L}\frac{z_0}{p_1}\sum_{k=0}^{\bar{N}-1}S_k \cr &&\hspace*{-0.2 cm} -\left(\rho_+-\frac{1}{L}\right)\frac{\frac{w_{21} X_1}{1-X_1}+\frac{ w_{12} X_2}{1-X_2}}{\frac{w_{21}}{1-X_1}+\frac{w_{12}}{1-X_2}}-\frac{1}{L}\frac{\frac{w_{21} Y_1}{1-Y_1}+\frac{w_{12} Y_2}{1-Y_2}}{\frac{w_{21}}{1-Y_1}+\frac{w_{12}}{1-Y_2}}, 
 \label{eq:2-point}
 \eea
\begin{figure}[t]
\centering \includegraphics[width=8.6 cm]{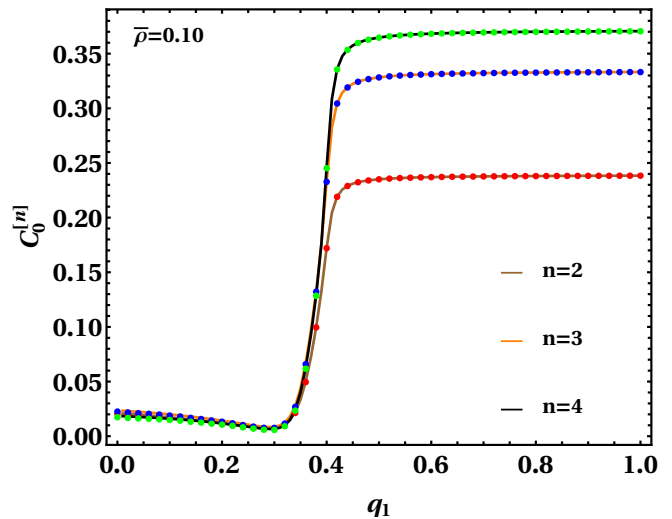}
\caption{The figure shows abrupt change in the $n$-point correlation between consecutive vacancies near the transition point $q_1=p_1$. For $q_1>p_1$, the growth of the correlation with increasing $n$ implies the tendency of vacancies to form larger cluster. Solid lines and dots correspond to analytical calculation and Monte Carlo simulation results respectively. The parameters used are $L=10^3, p_1=0.4, p_2=1.0, q_2=0.5, \epsilon=0.1, w_{12}=1.0, w_{21}=0.1, \rho_+=0.25, \rho_0=0.4$.}
\label{fig:correlation}
\end{figure}
 where $S_k$ is given by Eq.~(\ref{eq:pf6}) and $X_1 =\frac{z_0}{p_1}+\frac{z_0}{\epsilon}\frac{q_1}{p_1}, 
 X_2 = \frac{z_0}{p_2}+\frac{z_0}{\epsilon}\frac{q_2}{p_2}, 
 Y_1 = \frac{z_0}{p_1} S_{\bar{N}}, 
 Y_2 = \frac{z_0}{p_2}+\frac{z_0}{p_1}\frac{q_2}{p_2}S_{\bar{N}-1}$. The sum $\frac{1}{L}\sum_{k=0}^{\bar{N}-1}S_k$ in Eq.~(\ref{eq:2-point}), and consequently $C^{[2]}_0$, can be obtained in closed form, given by Eq.~(\ref{eq:sumsk}) in 
Appendix~\ref{app:mpa}. We also provide the successive expressions of $C^{[n]}_0$ for different $n$, along with the corresponding formula for any general $n$ in Eq.~(\ref{eq:allcorr})  (Appendix~\ref{app:mpa}). 
The variation of $C^{[n]}_0$ with the counter-flow parameter $q_1$ for different values of $n$, is presented in Fig.~\ref{fig:correlation}.  For each $n$ the correlation, similar to density and current, displays the existence of two different phases for $q_1<p_1$ and $q_1>p_1$. In Fig.~\ref{fig:correlation}, we observe a steep increase in the correlations near $q_1\lesssim p_1$. 
Interestingly, the correlations between consecutive vacancies increase considerably with increasing $n$ for $q_1>p_1$. This implies more number of vacancies prefer to stick together in the parameter regime $q_1>p_1$. It appears to be a direct evidence of macroscopic cluster formation of vacancies in the clustering phase. We would like to use the fact that $z_0$ remains almost zero in the clustering phase (inset of Fig.~\ref{fig:density}), in Eq.~(\ref{eq:2-point}). This simplifies the expression for $C^{[2]}_0$ considerably in the clustering phase and leads to
\be
C^{[2]}_0\approx \rho_0-\rho_0^2, \hspace*{0.4 cm}\mathrm{for}\,\,q_1>p_1.
\label{eq:2pc}
\ee
The above result is consistent with $C^{[2]}_0$ being constant for $q_1>p_1$ in Fig.~\ref{fig:correlation}, with the corresponding value agreeing to the same given in Eq.~(\ref{eq:2pc}). In fact, similar arguments apply for general $n$, and Eq.~(\ref{eq:allcorr}) consequently results in the following particularly simple expression for $C^{[n]}_0$ in the clustering phase
\be
C^{[n]}_0\approx \rho_0-\rho_0^n, \hspace*{0.4 cm}\mathrm{for}\,\,q_1>p_1.
\label{eq:npc}
\ee
We understand from Eq.~(\ref{eq:npc}) that with increasing $n$, $C^{[n]}_0$ approaches to $\rho_0$ which is simply the density of the vacancies, and this statement is also evident from Fig.~\ref{fig:correlation}.
\section{Effect of non-ergodicity on clustering}
\label{non-ergodic}
The discussion of the two different phases up to now, corresponds to the initial configuration Eq.~(\ref{eq:typeIII}). The features of the free flowing phase and the clustering phase would remain the same for initial configurations that can be prepared by permuting the positions of vacancies in Eq.~(\ref{eq:typeIII}). But, owing to the non-ergodic nature of $2$-ASEP-IAF, it is important to ask about the effect of the rearrangement of the two species and impurities in the initial configuration Eq.~(\ref{eq:typeIII}). To answer this, in this section we would like to investigate the effect of non-ergodicity on the clustering phenomena. We consider the following variation  of the initial configuration  Eq.~(\ref{eq:typeIII})
\begin{equation}
D_2 A..D_2A\,\,D_1D_1 A..D_1D_1 A\,\,D_1A..D_1 A\,\,D_1..D_1\,\,E..E. 
\label{eq:eta}
\end{equation}
\begin{figure}[t] 
  \centering \includegraphics[width=8.6 cm]{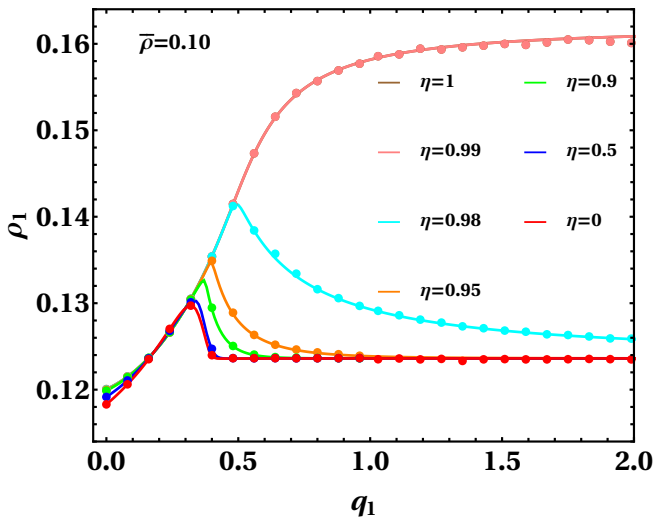}
  \caption{The figure demonstrates the effect of  rearrangement parameter $\eta$ (characterizing non-ergodicity) on the clustering phase. As $\eta$ decreases, the distinction between the free flowing phase and clustering phase becomes more and more abrupt. Particularly, the density remains monotonic for large $\eta\approx1$, whereas it is non-monotonic for small and moderate values of $\eta$. Solid lines and dots correspond to analytical calculation and Monte Carlo simulation results respectively. The parameters used are $L=10^3, p_1=0.4, p_2=1.0, q_2=0.5, \epsilon=0.1, w_{12}=1.0, w_{21}=0.1, \rho_+=0.25, \rho_0=0.4$.}
\label{fig:eta}
\end{figure}
We emphasize that all the input parameters $(p_{1,2},\epsilon, w_{12,21},\rho_0,\rho_+,\bar{\rho})$ are same both for Eqs.~(\ref{eq:typeIII}) and (\ref{eq:eta}). The difference between these two initial configurations lies in the {\it rearrangement} of the non-flipping species $1$ particles. In Eq.~(\ref{eq:eta}), we have two types of non-flipping species $1$ particles, {\it isolated} (left $D_1$ of unit $D_1D_1A$, cannot come in contact with another non-flipping $D_1$), and {\it non-isolated} (belongs to sequence $D_1..D_1$). For Eq.~(\ref{eq:typeIII}), we have only the non-isolated type. However, the total density of non-flipping $D_1$ is $\bar{\rho}$, same for both Eq.~(\ref{eq:typeIII}) and Eq.~(\ref{eq:eta}). To quantify their difference, we denote the fraction of isolated non-flipping species $1$ particles in Eq.~(\ref{eq:eta}) as $\eta$. 
If we denote the densities of isolated and non-isolated types of non-flipping species $1$ particles by $\rho_{\mathrm{iso}}$ and $\rho_{\mathrm{niso}}$ respectively, then $\eta$ is given by
\be
\eta= \frac{\rho_{\mathrm{iso}}}{\bar{\rho}}=1-\frac{\rho_{\mathrm{niso}}}{\bar{\rho}},
\label{eq:eta}
\ee
where $\bar{\rho}$ is defined through Eq.~(\ref{eq:rb}). The initial configuration in Eq.~(\ref{eq:typeIII}) corresponds to $\eta=0$.
Since the variation of $\eta$ simply rearranges the non-flipping $D_1$-s in the initial configuration, we denote it as {\it rearrangement parameter}. Thus $\eta$ appears as a hallmark of non-ergodicity. 
We present the behavior of average species density $\rho_1$ as a function of $\eta$ in Fig.~\ref{fig:eta},  obtained in Eq.~(\ref{eq:obs1}) (Appendix~\ref{app:mpa}), following similar methods discussed in Section~\ref{init}. With decreasing $\eta,$ the distinction between two phases become more evident in Fig.~\ref{fig:eta} and the fall of the density to constant value gets sharper. Intriguingly, for small or moderate values of rearrangement parameter, the density is non-monotonic, contrary to its monotonic nature for $\eta\approx1$.  Figure~\ref{fig:eta} thus illustrates that the onset of clustering strongly depends on $\eta$, and therefore on the choice of the initial configuration. In reality, often there are restrictions on the tunable range of the tuning parameter. For the variable range of the parameter under consideration (e.g. say $q_1\in(0,1)$ for some system), Fig.~\ref{fig:eta} tells us which initial configurations are more prone to clustering in the steady state and which initial configurations are suitable to avoid such clustering, for a given value of $q_1$. Taking into consideration the possible mapping of the present model to a narrow two-lane system with counter-flow, the analysis of diagrams like Fig.~\ref{fig:eta} might help in predicting the chances of jamming in the steady state starting from different initial configurations.
\section{Analysis of the case $q_1-p_1=p_2-q_2$: mapping to run-and-tumble particles}
\label{rtp}
\begin{figure}[t]
\centering
\subfigure[]{\includegraphics[scale=0.65]{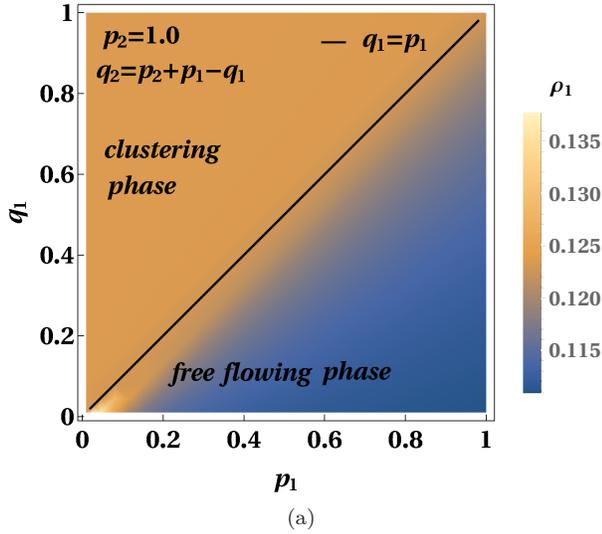}}\\
\subfigure[]{\includegraphics[scale=0.65]{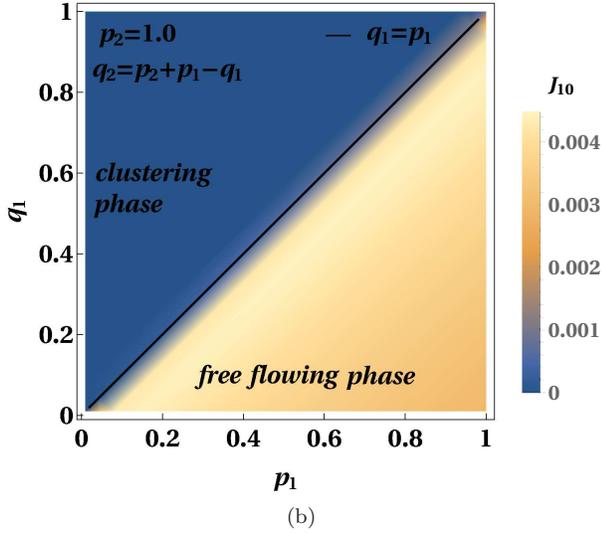}}\\
\caption{The figures (a) and (b) show heat-maps for density $\rho_1$ and current $J_{10}$ respectively, in the $p_1-q_1$ plane, for the special case $q_2-p_2=p_1-q_1$ that maps the $2$-ASEP-IAF to a system of RTPs. Both figures exhibit the existence of two different phases for $q_1>p_1$ and $q_1<p_1$. The current being close to zero for $q_1>p_1$ in figure (b) confirms that this parameter region corresponds to the clustering phase, while $q_1<p_1$ with non-zero current gives the free flowing phase.
The parameters used are $L=10^3, p_2=1.0, \epsilon=0.1, w_{12}=1.0, w_{21}=0.1, \rho_+=0.25, \rho_0=0.4$.}.
\label{fig:rtp_heatmap}
\end{figure}
\begin{figure}[t]
\centering
\subfigure[]{\includegraphics[scale=0.5]{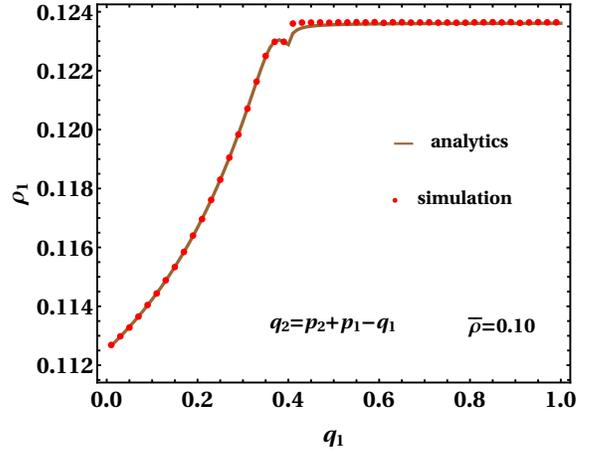}}\\
\subfigure[]{\includegraphics[scale=0.5]{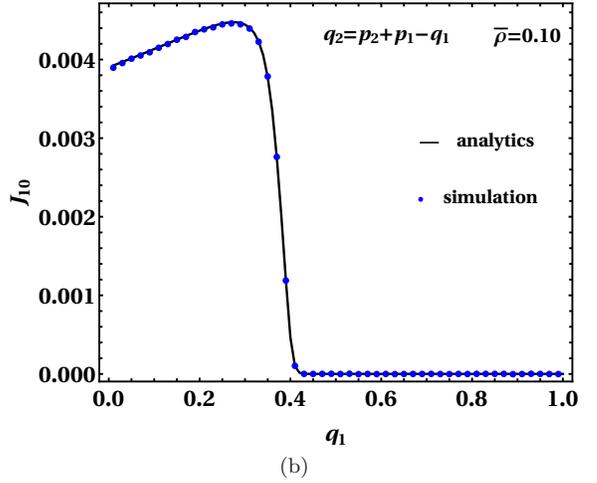}}\\
\caption{The figures (a) and (b) show two different phases for the RTPs, through the variation of average species density $\rho_1$ and average drift current $J_{10}$ respectively, as functions of the parameter $q_1$. For $q_1>p_1$ and $q_2<p_2$, $\rho_1$ is constant and $J_{10}$ is vanishingly small, characterizing the phase as clustering phase. For $q_1<p_1$ and $q_2>p_2$, we observe the free flowing phase where both density and current vary considerably with $q_1$. The solid lines and the dots correspond to analytical and Monte Carlo simulation results,  respectively. The parameters used are $L=10^3, p_1=0.4, p_2=1.0, \epsilon=0.1, w_{12}=1.0, w_{21}=0.1, \rho_+=0.25, \rho_0=0.4$.}.
\label{fig:rtprhoj}
\end{figure}
\begin{figure}[t]
\centering
\subfigure[]{\includegraphics[scale=0.5]{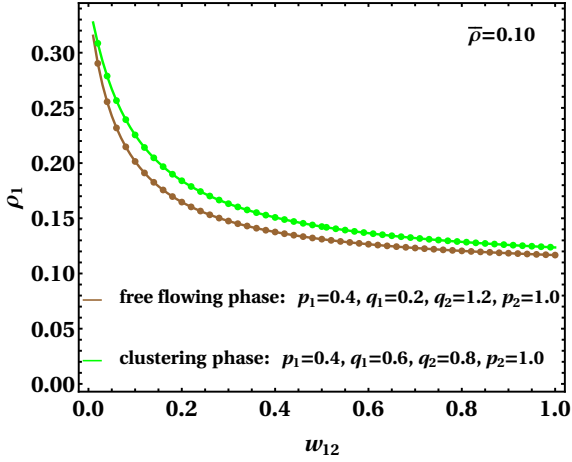}}\\
\subfigure[]{\includegraphics[scale=0.5]{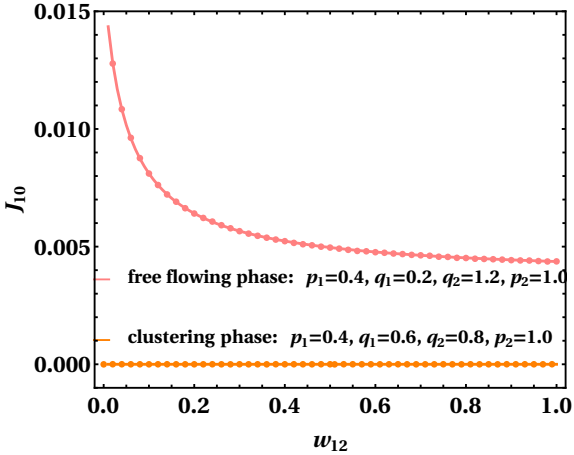}}\\
\caption{The figures (a) and (b) illustrate the effect of the tumbling rate $w_{12}$ on the two phases. The figure (a) exhibits similar qualitative behavior (monotonic decrease) of the density $\rho_1$ in both phases. The figure (b) shows that the clustering phase remains unaffected with the change of $w_{12}$, whereas the current decreases monotonically with $w_{12}$ in the free flowing phase. The solid lines and the dots correspond to analytical and Monte Carlo simulation results, respectively. The parameters used are $L=10^3, \epsilon=0.1, w_{21}=0.1, \rho_+=0.25, \rho_0=0.4$.}.
\label{fig:rtprhojw12}
\end{figure}
\begin{figure}[t]
\centering
\subfigure[]{\includegraphics[scale=0.5]{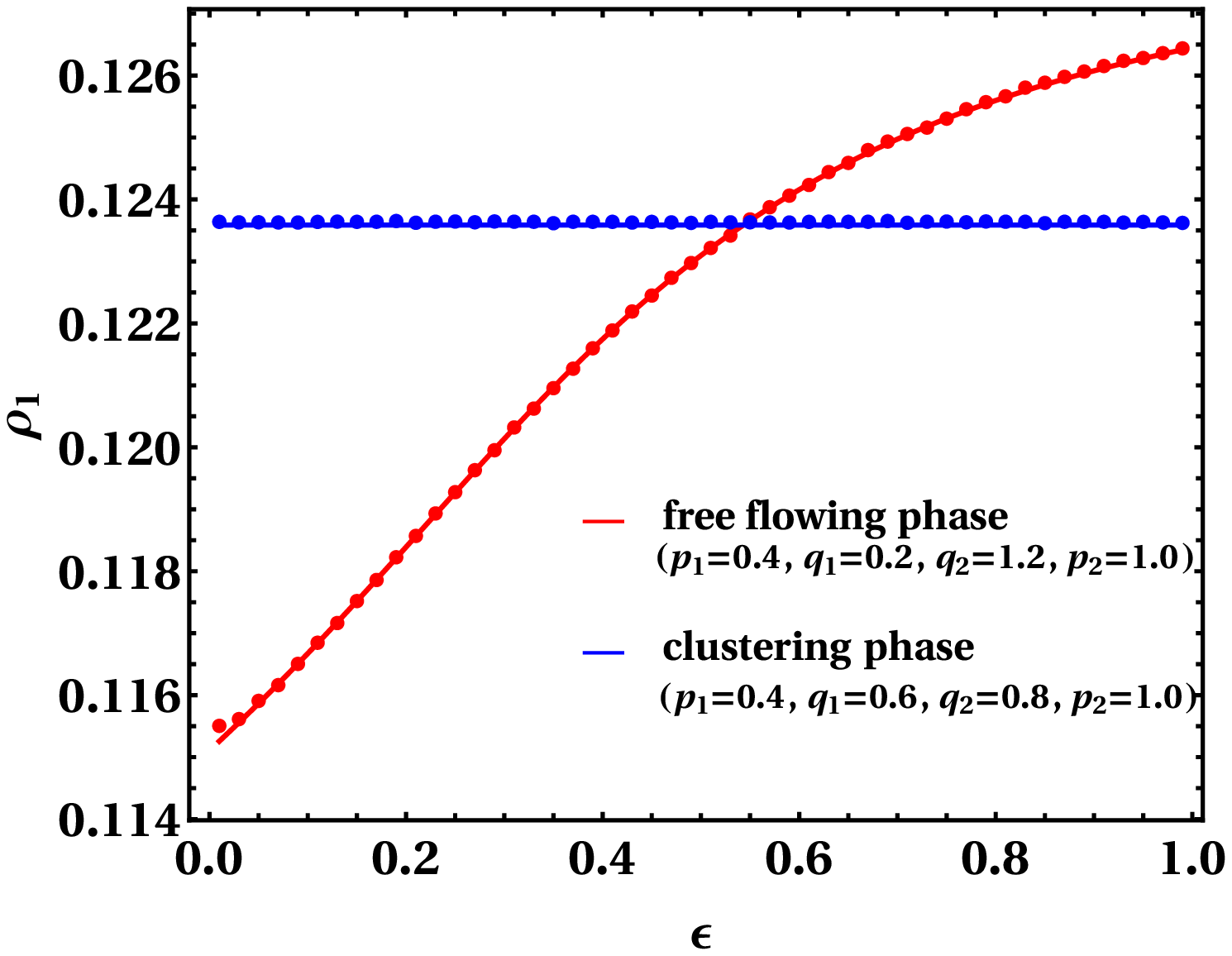}}\\
\subfigure[]{\includegraphics[scale=0.5]{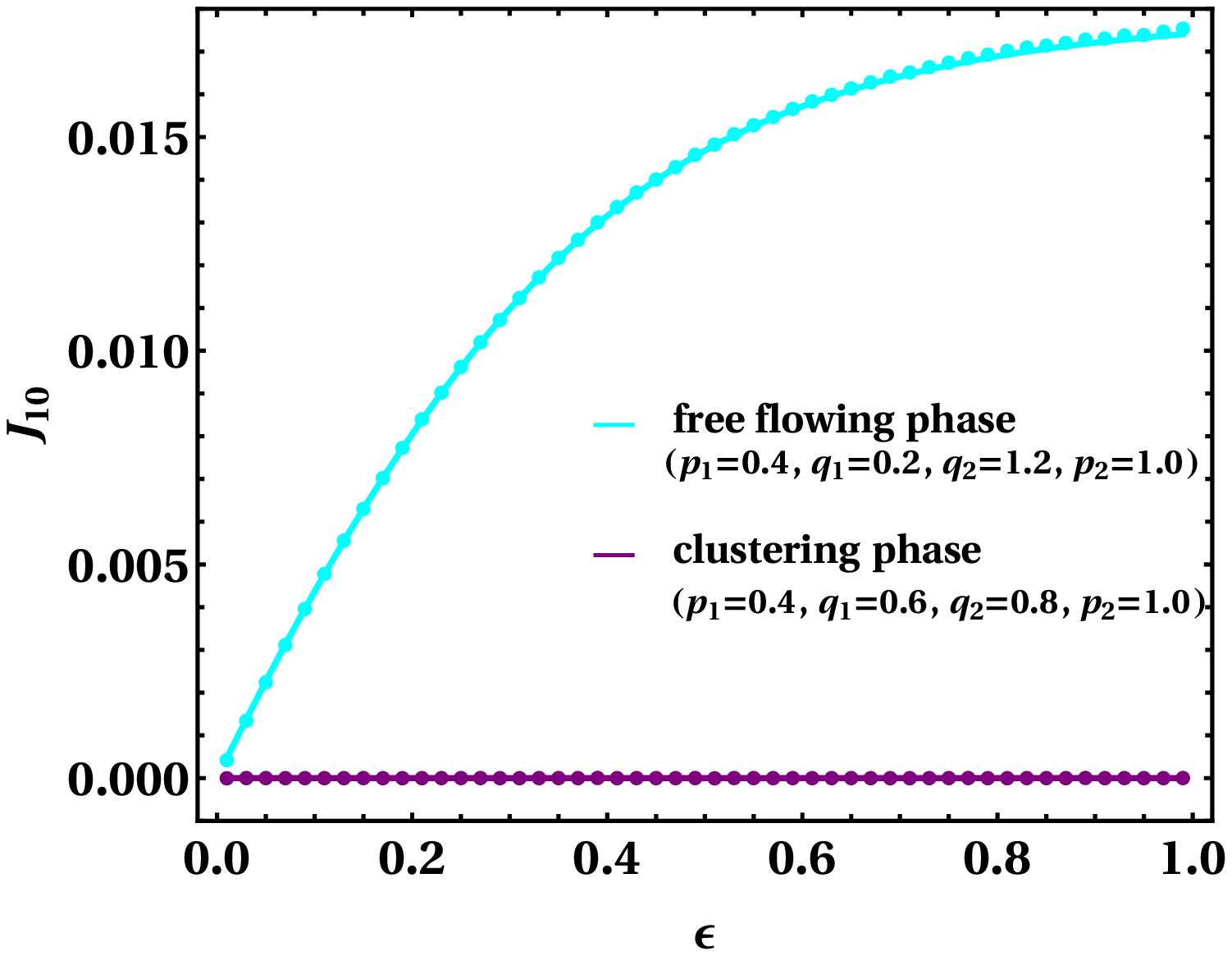}}\\
\caption{The figures (a) and (b) illustrate the effect of the drift $\epsilon$ of tumbling reagents (i.e. impurities) on the two phases. Both the density and current, in figures (a) and (b) respectively, show that the variation of $\epsilon$ has no effect in the clustering phase, whereas both the observables increase monotonically with $\epsilon$ in the free flowing phase. The solid lines and the dots correspond to analytical and Monte Carlo simulation results,  respectively. The parameters used are $L=10^3, w_{12}=1.0, w_{21}=0.1, \rho_+=0.25, \rho_0=0.4$.}.
\label{fig:rtprhojeps}
\end{figure}
In this section, we would like to consider the subspace of the parameter space that satisfies the constraint $q_1-p_1=p_2-q_2$. This special case is worthy of some detailed analysis because of its underlying connection to run-and- tumble particles. 

Self-propelled objects such as bacteria respond to chemical stimulus (e.g. nutrients or harmful substances) present in the environment through the process of {\it chemotaxis}. Chemotaxis, in a simple language, can be understood as run-and-tumble motion. {\it Run} corresponds to motion along fixed direction and {\it tumble} corresponds to intermediate random reorientation of direction of motion. Individual swimming flagellated bacteria e.g. E. coli in suspensions exhibit run-and-tumble dynamics   \cite{Turner_2000,Berg_2004,Sidortsov_2017,Beer_2019}. More interestingly,  there are run-and-tumble wild bacteria e.g. Vibrio ordalli in oceans, for which even the chemical stimulus can be dynamic \cite{Brumley_2019}. The creation, destruction and diffusion of the short-lived chemical stimulus in ocean create the dynamic chemical gradient  \cite{Brumley_2019}. Bacteria like E. coli or V. ordalli can accumulate near chemical stimulus to create clusters \cite{Turner_2000,Berg_2004,Brumley_2019}. Such run-and-tumble motion and clustering of bacteria can be best described in two dimensional continuous space, which poses difficulties for exact analysis. Rather, one dimensional discrete lattice models, although simpler adaptation of the complex bacterial systems, can capture the run-and-tumble dynamics and  cluster formation.  
In this connection, it is intriguing to note that the $2$-ASEP-IAF [Eq.~(\ref{eq:dynamics})] can be  mapped to a system of many RTPs, under the special circumstance $q_1-p_1=p_2-q_2$. 

In this case, the net bias of the two species are equal and opposite to each other, and therefore the two species can be considered as the manifestations of two possible orientations of RTPs in one dimension. The impurities act as origins of the tumbling process. The flip between species can be interpreted as tumbling of directions. Particularly, when the impurity causes flip to a species that has net bias in the direction same as that of the impurities, the impurity at that instant can be thought of a attracting reagent or nutrients. On the other hand, when the resulting species from flip has net bias opposite to the impurity bias, the impurity at that instant is considered as a repelling reagent for tumbling process. Note that apart from the RTPs (constituted by the $D_1$-s and $D_2$-s in the sequences of $\mathcal{Y}=D_1A, D_2A$ in Eq.~(\ref{eq:typeIII})) and sources of tumbling process (i.e. impurities $A$ in the sequences of $\mathcal{Y}=D_1A, D_2A$ in Eq.~(\ref{eq:typeIII})), there are particles which do not tumble in the sense that their direction of net bias remain unchanged (the non-flipping $D_1$-s in the sequence of $\mathcal{Y}=D_1$ in Eq.~(\ref{eq:typeIII})). With this mapping of the $2$-ASEP-IAF to RTPs, we would like to discuss the behaviors of the average density [Eq.~(\ref{eq:rho})] and current [Eq.~(\ref{eq:j})] of the species for the special case $q_1-p_1=p_2-q_2$, as a function of the system parameters.

We present the heat-maps of the average density $\rho_1$ and current $J_{10}$ in Figs.~\ref{fig:rtp_heatmap}(a) and (b), in the $p_1-q_1$ plane. It should be mentioned that the mapping discussed in this section forces us to consider $q_2-p_2=p_1-q_1$, i.e. Fig.~\ref{fig:rtp_heatmap} strictly corresponds to the counter-flow situation (both for $q_1<p_1$ and $q_1>p_1$). This is contrary to the setup in Fig.~\ref{fig:heatmap} where we have natural flow for $q_1<p_1$ and counter-flow for $q_1>p_1$. In spite of this, we observe from Fig.~\ref{fig:rtp_heatmap} that, even within the counter-flow scenario, two different phases emerge showing distinct features for $q_1<p_1$ and $q_1>p_1$ that are qualitatively similar to Fig.~\ref{fig:heatmap}. For further analysis,  
in Figs.~\ref{fig:rtprhoj}(a) and (b), we present cross sections of Figs.~\ref{fig:rtp_heatmap}(a) and (b), respectively, at fixed $p_1$. Both Figs.~\ref{fig:rtprhoj}(a) and (b) exhibit two different phases, the free flowing phase and the clustering phase. For $q_1<p_1$ and $q_2>p_2$, both the density and current vary non-monotonically with $q_1$, which we understand to be the free flowing phase. Whereas, for $q_1>p_1$ and $q_2<p_2$, both $\rho_1$ and $J_{10}$ remain constant, specifically the current is vanishingly small identifying the phase as the clustering phase. In comparison to Fig.~\ref{fig:density} where $\rho_1$ attains maximum value in the free flowing phase, the maximum density for Fig.~\ref{fig:rtprhoj}(a) is achieved in the clustering phase. The cluster formation owes to the cumulative effect emerging from the presence of the non-tumbling sequence of $\mathcal{Y}=D_1$ in the initial configuration Eq.~(\ref{eq:typeIII}). We should mention that there is disagreement to some extent between our exact results and the Monte Carlo simulation results near $q_1=p_1$ in the clustering phase observed in Fig.~\ref{fig:rtprhoj}(a), whereas both results match well for all other values of $q_1$. This discrepancy, as per our current understanding, is due to the restriction of insufficient ensemble average near the transition point.  

It is quite natural to ask the effect of tumbling rates ($w_{12}$ and $w_{21}$) on the two phases. To investigate this, we present the variation of $\rho_1$ and $J_{10}$ as functions of the tumbling rate $w_{12}$ in Figs.~\ref{fig:rtprhojw12}(a) and (b), respectively, using our analytical findings (solid lines) from Eqs.~(\ref{eq:rho}) and (\ref{eq:j}), supported by Monte Carlo simulation results (dots). In each figure, we consider two situations, one corresponding to $q_1<p_1$ and $q_2>p_2$, and the other to $q_1>p_1$ and $q_2<p_2$. Particularly, Fig.~\ref{fig:rtprhojw12}(b) clearly shows the current remains vanishingly small for any value of $w_{12}$, in the case $q_1>p_1$ and $q_2<p_2$, suggesting that the formed cluster is stable to change of the tumbling rates. On the other hand, for $q_1<p_1$ and $q_2>p_2$, the current of species $1$ decreases monotonically with increasing tumbling rate $w_{12}$ (species $1$ to species $2$) as expected. However, the fact that the change of tumbling rate does not have any effect on the formed clusters, is not visible from the density characteristics in Fig.~\ref{fig:rtprhojw12}(a) which shows similar qualitative nature for both the free flowing and clustering phase. We would also like to explore the effect of the drift process of the impurities, that act as attracting reagent (nutrient) or repelling reagent causing the tumbles, on the two phases. The observables $\rho_1$ and $J_{10}$ in Figs. \ref{fig:rtprhojeps}(a) and (b), respectively, exhibit that the clustering phase is undisturbed by the change of the impurity drift rate $\epsilon$, where as both the average density and current increases with increasing $\epsilon$ in the free flowing phase. Thus we have shown that, starting from initial configuration Eq.~(\ref{eq:typeIII}), the system of RTPs (along with attracting or repelling reagents causing tumbles) can transit between free flowing phase and clustering phase, with the variation of the control parameter $q_1$ that tunes the net biases of the two tumbling directions. Interestingly, the formed clusters are stable against any change of the tumbling rates $w_{12}$ (and $w_{21}$) and reagent (i.e. impurity causing tumble) drift rate $\epsilon$.   
\section{Summary}
\label{summary}
We have demonstrated the formation of clusters induced by counter-flow in a non-ergodic system. To illustrate this, we have considered two species asymmetric simple exclusion process along with impurities. Apart from the drift of the species and impurities, additionally the impurities activate flips between the two species. The exact analytical results obtained for the observables (average species density, drift current, spatial correlation),  show two distinct phases, {\it free flowing phase} and {\it clustering phase}, as functions of the counter-flow parameter $q_1$, which basically controls the flow direction of species $1$ by tuning its net bias.  
The {\it free flowing phase} is specified by {\it non-monotonic density} and {\it non-monotonic finite current}, whereas the {\it clustering phase} is characterized by {\it constant density} and {\it vanishing current}. The clustering situation can be thought as an equivalent system almost devoid of vacancies, where the drift dynamics is {\it almost} absent. This is compatible with  the vanishing current and the constant density in the clustering phase. Specifically, the growth of $n$-point spatial  correlation between $n$ consecutive vacancies with increasing $n$, directly implies the accumulation of vacancies to form one macroscopic cluster, along with another macroscopic cluster formed by all the particles.  The heat maps of density and current in $p_1-q_1$ plane show that the two phases are demarcated by the line $q_1=p_1$ ($p_1$ and $q_1$ being right and left hop rates of species $1$, respectively). This is further clarified by the sharp descents of density and current and the sharp ascent in spatial correlation between vacancies near $q_1=p_1$.
The slow drift of the cluster in the clustering phase can be observed by tracking individual configurations over long time. The effect of non-ergodicity on the system is characterized through a rearrangement parameter, which only permutes the positions of some particles while keeping all the input parameters fixed. We observe that the choice of initial configuration affects the onset of clustering significantly. In fact, for certain initial configurations, we do not see signatures of clustering for finite values of $q_1$  and the corresponding densities increase monotonically in contrast to the non-monotonic densities for initial configurations showing cluster formations. Interestingly, for a special case of the microscopic dynamics when the net bias of the two species are equal and opposite to each other, we map the $2$-ASEP-IAF to a system of RTPs. The species with net bias to right (left) can be considered as right (left) running RTPs, where the impurities act as tumbling reagents that cause the tumbling of RTPs, equivalent to the flip process in $2$-ASEP-IAF. Notably, although this mapping is valid only in the counter-flow situation, we still observe two different phases, the free flowing phase and the clustering phase. We further find that the clustering phase remains stable against the variation of the tumbling rate and drift of the tumbling reagents. We believe that our analysis supported by exact analytical results enlightens the understanding of clustering phenomena. 

The model studied here, having resemblance to two-lane traffics and RTPs in active matter, points towards analytical understanding of traffic jams and clustering active matter systems. Further careful and thorough   investigations are required to establish such connections. It would be interesting to study variations of the local microscopic dynamics considered here, that can produce dynamical ways to get rid of the clustering phase. It would be important to explore the effects of boundary conditions, e.g. allowing entries and exits of selective particles or all particles, on the different phases of the system.

{\it Acknowledgements.-} This work is partially supported by the Grants-in-Aid
for Scientific Research (JSPS KAKENHI Grant No. JP21H01006). A.K.C. gratefully acknowledges postdoctoral fellowship from the YITP.    The numerical calculations have been done on the cluster Yukawa-21 at the YITP. 

\appendix
\section{Derivation of observables}
\label{app:mpa}
In this appendix, we sketch the steps for calculating  the average species densities and $n$-point correlation function between consecutive vacancies. These steps essentially follow the methods discussed in obtaining the partition function in Section~\ref{init}.  
The average density $\rho_1$ of species $1$, can be written as
\begin{widetext}
\begin{eqnarray}
\rho_1= \langle 1 \rangle &=& \bar{\rho}+  \frac{\left(\rho_+-\frac{1}{L}\right)}{Q} \sum_{m_1=0}^{\infty}\dots \sum_{n_{\bar{N}}=0}^{\infty}\,\, \mathrm{Tr}\left[D_1(z_0 E)^{m_1}A(z_0 E)^{\bar{m}_1}\,\prod_{i=2}^{N_+}(D_1+D_2)(z_0 E)^{m_i}A(z_0 E)^{\bar{m}_i}\,\,\prod_{j=1}^{\bar{N}}D_1 (z_0 E)^{n_j}\right] \cr && + \frac{1}{L} \frac{1}{Q} \sum_{m_1=0}^{\infty}\dots \sum_{n_{\bar{N}}=0}^{\infty}\,\, \mathrm{Tr}\left[\prod_{i=1}^{N_+-1}(D_1+D_2)(z_0 E)^{m_i}A(z_0 E)^{\bar{m}_i}\,\,D_1(z_0 E)^{m_{N_+}}A(z_0 E)^{\bar{m}_{N_+}}\,\,\prod_{j=1}^{\bar{N}}D_1 (z_0 E)^{n_j}\right].\nonumber \\
\label{eq:ob1}
\end{eqnarray}
\end{widetext}
The term $\bar{\rho}$ appears directly due to the initial density $\bar{\rho}$ of the non-flipping species $1$ particles. In the second part with pre-factor $\left(\rho_+-\frac{1}{L}\right)$, we place at least one $D_1$ in  a flipping term whereas any other flipping term can have $D_1$ or $D_2$, the density of such terms is $\left(\rho_+-\frac{1}{L}\right)$. The last part contributes due to the $D_1$ that is placed in the last flipping term after which the non-flipping $D_1$-s start. To proceed, we would use the explicit matrix representations in Eq.~(\ref{eq:matrices}) and the expressions derived in Eqs.~(\ref{eq:pf3}) and (\ref{eq:pf4}) to evaluate the matrix strings in Eq.~(\ref{eq:ob1}). {This leads us to the final formula for the average species density in Eq.~(\ref{eq:rho}).}
We can follow similar procedure to calculate the average species densities for the general initial configuration Eq.~(\ref{eq:eta}) with nonzero $\eta$ in Section~\ref{non-ergodic}. In this case, we get,
\begin{widetext}
\begin{eqnarray}
\rho_1 (\eta)&=&\bar{\rho}+\left(\rho_+-\frac{1}{L}-\eta\bar{\rho}\right)\frac{\frac{w_{21}}{1-\frac{z_0}{p_1}S_0}}{\left[\frac{w_{21}}{1-\frac{z_0}{p_1}S_0}+\frac{w_{12}}{1-\frac{z_0}{p_2}\left(1+\frac{q_2}{p_1} S_{-1}\right)}\right]} + \eta\bar{\rho} \frac{\frac{w_{21}}{1-\frac{z_0}{p_1}S_1}}{\left[\frac{w_{21}}{1-\frac{z_0}{p_1}S_1}+\frac{w_{12}}{1-\frac{z_0}{p_2}\left(1+\frac{q_2}{p_1} S_{0}\right)}\right]}\cr && +\frac{1}{L} \frac{\frac{w_{21}}{1-\frac{z_0}{p_1}S_{(1-\eta)\bar{N}}}}{\left[\frac{w_{21}}{1-\frac{z_0}{p_1}S_{(1-\eta)\bar{N}}}+\frac{w_{12}}{1-\frac{z_0}{p_2}\left(1+\frac{q_2}{p_1} S_{(1-\eta)\bar{N}-1}\right)}\right]},\cr
\rho_2 (\eta)&=& \bar{\rho}+\rho_+-\rho_1(\eta),
\label{eq:obs1}
\end{eqnarray}
\end{widetext}
It is straightforward to check that $\eta=0$ in Eq.~(\ref{eq:obs1}) gives the density in Eq.~(\ref{eq:rho}).

The formal expression for two-point nearest neighbor correlation between vacancies is
\begin{equation}
C^{[2]}_0=\langle 00 \rangle -\rho_0^2.
 \label{eq:c00}
\end{equation}
It is difficult to calculate $\langle 00 \rangle$ directly using the matrix representations, rather it is easier to use the following conservation 
\begin{equation}
\langle 00 \rangle +\langle 01 \rangle +\langle 02 \rangle +\langle 0+ \rangle =\rho_0.
 \label{eq:cons}
\end{equation}
Using Eq.~(\ref{eq:cons}) into Eq.~(\ref{eq:c00}), we get
\begin{equation}
C^{[2]}_0=\rho_0 -\rho_0^2-\langle 01 \rangle -\langle 02 \rangle -\langle 0+ \rangle.
 \label{eq:c00_1}
 \end{equation}
 Evaluating $\langle 01 \rangle$, $\langle 02 \rangle$ and $\langle 0+ \rangle$ using the matrix representations, we finally obtain the following expression for $C_{00}$ as
 \begin{eqnarray}
&& \hspace*{-0.2 cm} C^{[2]}_0=\rho_0 -\rho_0^2-\left(\rho_+-\frac{1}{L}\right)\frac{z_0}{\epsilon}-\frac{1}{L}\frac{z_0}{p_1}\sum_{k=0}^{\bar{N}-1}S_k \cr &&\hspace*{-0.2 cm} -\left(\rho_+-\frac{1}{L}\right)\frac{\frac{w_{21} X_1}{1-X_1}+\frac{ w_{12} X_2}{1-X_2}}{\frac{w_{21}}{1-X_1}+\frac{w_{12}}{1-X_2}}-\frac{1}{L}\frac{\frac{w_{21} Y_1}{1-Y_1}+\frac{w_{12} Y_2}{1-Y_2}}{\frac{w_{21}}{1-Y_1}+\frac{w_{12}}{1-Y_2}}, \nonumber\\
 \label{eq:c00_final}
 \end{eqnarray}
 where
 \begin{eqnarray}
 X_1 &=& \frac{z_0}{p_1}+\frac{z_0}{\epsilon}\frac{q_1}{p_1}, \cr
 X_2 &=& \frac{z_0}{p_2}+\frac{z_0}{\epsilon}\frac{q_2}{p_2}, \cr
 Y_1 &=& \frac{z_0}{p_1} S_{\bar{N}}, \cr
 Y_2 &=& \frac{z_0}{p_2}+\frac{z_0}{p_1}\frac{q_2}{p_2}S_{\bar{N}-1}.
  \label{eq:all}
 \end{eqnarray}
Clearly, the two-point correlation would have a closed form if the sum $\sum_{k=0}^{\bar{N}-1}S_k$ has a closed form, and indeed this can be evaluated as
\begin{eqnarray}
&&\frac{1}{L}\sum_{k=0}^{\bar{N}-1}S_k 
= \bar{\rho} \frac{p_1}{(p_1-q_1)}\cr &&+\frac{1}{L}\left[1-\left(\frac{q_1}{p_1}\right)^{\bar{N}}\right]\left(\frac{p_1-q_1}{\epsilon}-1\right)\frac{p_1q_1}{(p_1-q_1)^2}.
 \label{eq:sumsk}
\end{eqnarray}
So, we have evaluated the two-point correlation function $C^{[2]}_0$ exactly. Of course, $z_0$ has to be calculated from the density-fugacity relation. Importantly in Eq.~(\ref{eq:sumsk}), note that we could scale $\bar{N}$ by system-size $L$ properly in the first term so that it becomes a function of $\bar{\rho}$, but this is not possible in case of the second term where $\bar{N}$ appears in the power. 

In fact, using the result of $C^{[2]}_0$, we can calculate $C^{[3]}_0$ and then $C^{[4]}_0$ using $C^{[3]}_0$, and so on. For simplified notations, we denote $C_{0}^{[n]}$ as the correlation between consecutive $n$ vacancies [Eq.~(\ref{eq:n-point})], then we obtain in iterative way,
\begin{widetext}
\begin{eqnarray}
 C_{0}^{[2]}&=& \rho_0 -\rho_0^2-\left(\rho_+-\frac{1}{L}\right)\frac{z_0}{\epsilon}-\frac{1}{L}\frac{z_0}{p_1}\sum_{k=0}^{\bar{N}-1}S_k  -\left(\rho_+-\frac{1}{L}\right)\frac{\frac{w_{21} X_1}{1-X_1}+\frac{ w_{12} X_2}{1-X_2}}{\frac{w_{21}}{1-X_1}+\frac{w_{12}}{1-X_2}}-\frac{1}{L}\frac{\frac{w_{21} Y_1}{1-Y_1}+\frac{w_{12} Y_2}{1-Y_2}}{\frac{w_{21}}{1-Y_1}+\frac{w_{12}}{1-Y_2}}, \cr \cr
 C_{0}^{[3]}&=& { C_{0}^{[2]}+\rho_0^2 -\rho_0^3}-\left(\rho_+-\frac{1}{L}\right)\left(\frac{z_0}{\epsilon}\right)^{ 2}-\frac{1}{L}\left(\frac{z_0}{p_1}\right)^{ 2}\sum_{k=0}^{\bar{N}-1}\left(S_k\right)^{ 2}  -\left(\rho_+-\frac{1}{L}\right)\frac{\frac{w_{21} { X_1^2}}{1-X_1}+\frac{ w_{12} { X_2^ 2}}{1-X_2}}{\frac{w_{21}}{1-X_1}+\frac{w_{12}}{1-X_2}}-\frac{1}{L}\frac{\frac{w_{21} { Y_1^2}}{1-Y_1}+\frac{w_{12} { Y_2^2}}{1-Y_2}}{\frac{w_{21}}{1-Y_1}+\frac{w_{12}}{1-Y_2}}, \cr \cr
C_{0}^{[4]}&=& { C_{0}^{[3]}+\rho_0^3 -\rho_0^4}-\left(\rho_+-\frac{1}{L}\right)\left(\frac{z_0}{\epsilon}\right)^{ 3}-\frac{1}{L}\left(\frac{z_0}{p_1}\right)^{ 3}\sum_{k=0}^{\bar{N}-1}\left(S_k\right)^{ 3}  -\left(\rho_+-\frac{1}{L}\right)\frac{\frac{w_{21} { X_1^3}}{1-X_1}+\frac{ w_{12} { X_2^ 3}}{1-X_2}}{\frac{w_{21}}{1-X_1}+\frac{w_{12}}{1-X_2}}-\frac{1}{L}\frac{\frac{w_{21} { Y_1^3}}{1-Y_1}+\frac{w_{12} {Y_2^3}}{1-Y_2}}{\frac{w_{21}}{1-Y_1}+\frac{w_{12}}{1-Y_2}}, \cr \cr
\dots &=& \dots \cr \cr
C_{0}^{[n]}&=& { C_{0}^{[n-1]}+\rho_0^{n-1} -\rho_0^n}-\left(\rho_+-\frac{1}{L}\right)\left(\frac{z_0}{\epsilon}\right)^{ n-1}-\frac{1}{L}\left(\frac{z_0}{p_1}\right)^{ n-1}\sum_{k=0}^{\bar{N}-1}\left(S_k\right)^{ n-1} \cr && -\left(\rho_+-\frac{1}{L}\right)\frac{\frac{w_{21} { X_1^{n-1}}}{1-X_1}+\frac{ w_{12} { X_2^ {n-1}}}{1-X_2}}{\frac{w_{21}}{1-X_1}+\frac{w_{12}}{1-X_2}}-\frac{1}{L}\frac{\frac{w_{21} { Y_1^{n-1}}}{1-Y_1}+\frac{w_{12} { Y_2^{n-1}}}{1-Y_2}}{\frac{w_{21}}{1-Y_1}+\frac{w_{12}}{1-Y_2}}.
\label{eq:allcorr}
\end{eqnarray}
\end{widetext}

Thus we have obtained the analytical formulae for average species densities, drift currents and $n$-point correlation between consecutive vacancies. The results in Eq.~(\ref{eq:allcorr}) have been used to present the behavior of the correlations with the variation of the counter flow parameter $q_1$ in Fig.~\ref{fig:correlation}.

\section{Density-fugacity relation: solution for special cases}
\label{app:rhoz}
Here we would like to state the explicit form of the density-fugacity relation, calculated from the partition function in Eq.~(\ref{eq:pf5}). This relation is used to solve the fugacity $z_0$ as a function of the input parameters. Consequently, we can replace the corresponding value of $z_0$ in the expressions of the observables e.g. in Eqs.~(\ref{eq:rho}), (\ref{eq:j}) and (\ref{eq:2-point}) so that they become functions of the input parameters only. The formal expression for the density-fugacity relation is given in Eq.~(\ref{eq:rhozapp}). Using Eq.~(\ref{eq:pf5}) in Eq.~(\ref{eq:rhozapp}), we have the following explicit form of the density-fugacity relation to solve,
\begin{widetext}
\begin{eqnarray}
&&  \frac{\rho_+}{1-\frac{z_0}{\epsilon}}+\frac{z_0\left(\rho_+-\frac{1}{L}\right)}{w_{21}(1-z_0 X'_2)+w_{12}(1-z_0 X'_1)}\left[w_{21}X'_1\frac{1-z_0X'_2}{1-z_0 X'_1}+w_{12}X'_2\frac{1-z_0X'_1}{1-z_0 X'_2}\right]\cr && +\frac{\frac{z_0}{L}}{w_{21}(1-z_0 Y'_2)+w_{12}(1-z_0 Y'_1)}\left[w_{21}Y'_1\frac{1-z_0Y'_2}{1-z_0 Y'_1}+w_{12}Y'_2\frac{1-z_0Y'_1}{1-z_0 Y'_2}\right]  +\frac{1}{L} \sum_{k=1}^{\bar{N}}\frac{1}{1-\frac{z_0}{p_1}S_{k-1}}
=\rho_0+\rho_++\bar{\rho},
 \label{eq:srhozexp}
\end{eqnarray}
\end{widetext}
with $X'_{1,2}=X_{1,2}/z_0$ and $Y'_{1,2}=Y_{1,2}/z_0$, where $X_{1,2}$ and $Y_{1,2}$ follow Eq.~(\ref{eq:all}), $S_k$ is given by Eq.~(\ref{eq:pf6}). The reason behind such rescaling by $z_0$, is simply to express $X'_{1,2}$ and $Y'_{1,2}$ as functions of the input parameters only. In general, we have solved Eq.~(\ref{eq:srhozexp}) in Mathematica to get $z_0$ for a given set of input parameters. Also note that the complexity of the equation increases with increasing number of non-flipping species $1$ particles $\bar{N}$, in terms of the highest degree of $z$ present in the polynomial of $z_0$ in  Eq.~(\ref{eq:srhozexp}). For some special cases with specific choices of the hop-rates, one can obtain closed form solutions for the fugacity $z_0$.

A particularly simple case corresponds to $\bar{N}=0$. This also implies $Y'_1=X'_1$ and $Y'_2=X'_2$, which can be seen directly from  
Eq.~(\ref{eq:all}) with the help of Eq.~(\ref{eq:pf6}). We further consider the special situation of $X'_1=X'_2$. The density-fugacity relation Eq.~(\ref{eq:srhozexp}) simplifies to
\begin{equation}
\frac{z_0 \rho_+ X'_1}{1-z_0 X'_1}+ \frac{\rho_+}{1-\frac{z_0}{\epsilon}}=\rho_0+\rho_+.
 \label{eq:srhozexp1}
\end{equation}
The above equation has the following solution
\begin{widetext}
\begin{equation}
z_0=\frac{(\rho_0+\rho_+)(1+\epsilon X'_1)+\sqrt{(\rho_0+\rho_+)^2(1+\epsilon X'_1)^2-4\rho_0\epsilon X'_1(\rho_0+2\rho_+)}}{2(\rho_0+2\rho_+)X'_1}.
 \label{eq:z0}
\end{equation}
\end{widetext}
To better understand the constraint on the hop rates for which we have got the exact solution Eq.~(\ref{eq:z0}), we explore the situation $X'_1=X'_2$, which basically boils down to
\begin{equation}
\epsilon=\frac{q_2 p_1-q_1p_2}{p_2-p_1}.
 \label{eq:eps}
\end{equation}
The above subspace of hop rates can create both natural flow and counter-flow situations and also includes the very special case $p_1=p_2$ and $q_1=q_2$. We should mention that, even without the assumption $X'_1=X'_2$, we have a quartic equation in $z_0$ that can be solved exactly in Mathematica. However, the solution of $z_0$ in that case is too lengthy to include here. 

Another noteworthy point is, the fugacity $z_0$ actually equals to the cluster velocity $v_{\mathrm{cl}}$ discussed in the main text (see inset of Fig.~\ref{fig:cj}). This is evident from the current-density relation Eq.~(\ref{eq:j}), which can be considered as $J=v_{\mathrm{cl}} \rho$ in the clustering phase. In fact, one can check the inset of Fig.~\ref{fig:cj} in the main text, obtained from Monte Carlo simulations, can be reproduced by calculating $z_0$ for the corresponding set of input parameters.

\section{A comment regarding the initial configuration}
\label{app:init}
We have considered step-like initial configuration [Eq.~(\ref{eq:typeIII})] in the main text. To elaborate, initially all the particles (both species and impurities) occupy consecutive lattice sites with no vacancy between them. Starting from such step-like initial configuration, in the free flowing phase, the vacancies get randomly distributed between the particles. On the other hand, in the clustering phase, any steady state configuration remains step-like, with the particle cluster shifting slowly to right. 

Naturally, the question arises if the cluster can be formed from an initial configuration which is not step-like, rather there are vacancies distributed between particles. The answer is {\it yes}. If we start from a non-step-like initial configuration given below,
\begin{widetext}
\begin{equation}
C(0)=D_2 E^{m_1} A E^{\bar{m}_1}\dots D_2 E^{m_{N_+/2}} A E^{\bar{m}_{N_+/2}}\,\,D_1 E^{n_1} A E^{\bar{n}_1}\dots D_1 E^{n_{N_+/2}} A E^{\bar{n}_{N_+/2}}\,\,D_1 E^{r_1} \dots D_1 E^{r_{\bar{N}}},
\label{eq:non-step_init}
\end{equation}
\end{widetext}
where the total number of vacancies is 
\begin{equation}
\sum_{i=1}^{N_+/2} \left(m_i+\bar{m}_i+n_i+\bar{n}_i\right)+ \sum_{j=1}^{\bar{N}} r_k=N_0.
\label{eq:rho0} 
\end{equation}
\begin{figure}[t]
  \centering \includegraphics[width=8.6 cm]{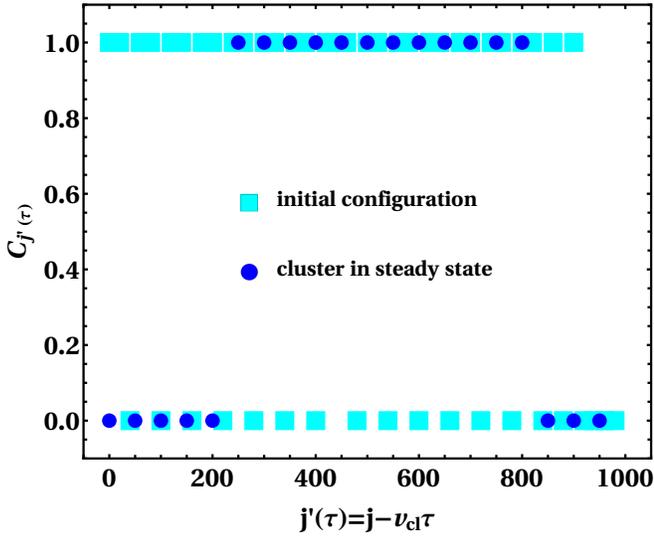}
  \caption{The figure demonstrates the step-like cluster  formation in the steady state, starting from a non-step-like initial configuration where vacancies are randomly distributed among particles. Here we consider $\tau=0$ as  the observation time of cluster in steady state, whereas the initial configuration corresponds to time $t=0$. The parameters used are $L=10^3, p_1=0.4, q_1=0.5, p_2=1.0, q_2=0.5, \epsilon=0.1, w_{12}=1.0, w_{21}=0.1, \rho_+=0.29, \rho_0=0.4$.}
\label{fig:non-step_init}
\end{figure}
Note that, the above initial configuration contains exact same ordering of particles just like the one [Eq.~(\ref{eq:typeIII})] studied in the main text and the total number of vacancies are also the same for both configurations, only difference being the initial configuration [Eq.~(\ref{eq:typeIII})] in the main text is step-like, whereas Eq.~(\ref{eq:non-step_init}) is non-step-like. Since the ordering of vacancies actually do not matter, both of these initial configurations lead to the same configuration sub-space in the steady state. Thereby all the characteristics of the system in the steady state remain same for both of these initial configurations. Thus we expect to see the clustering phenomena starting from initial configuration Eq.~(\ref{eq:non-step_init}) just like we did for the one in the main text (see Fig.~\ref{fig:cj}).  Indeed, in Fig.~\ref{fig:non-step_init}, we observe that the macroscopic cluster is formed in the steady state, while the initial configuration is non-step-like. So, the only advantageous and satisfactory thing about Eq.~(\ref{eq:non-step_init}), is the fact that the cluster is formed in a dynamic way from non-step-like initial configuration.
\section{Variation of observables with system size $L$}
\label{app:L}
In this appendix, we present the behaviors of average species density and drift current as function of $q_1$, for different system sizes $L$.
\begin{figure}[t]
\centering
\subfigure[]{\includegraphics[scale=0.55]{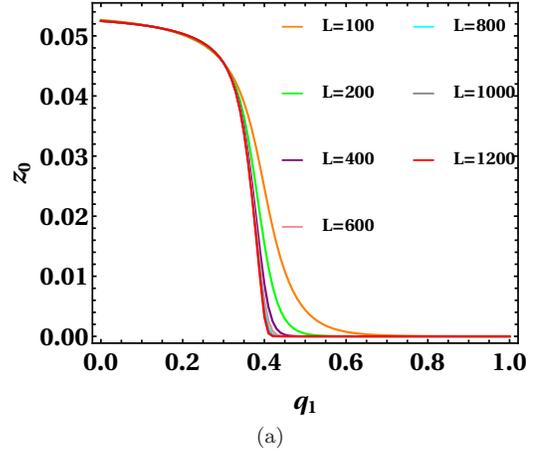}}\\
\subfigure[]{\includegraphics[scale=0.55]{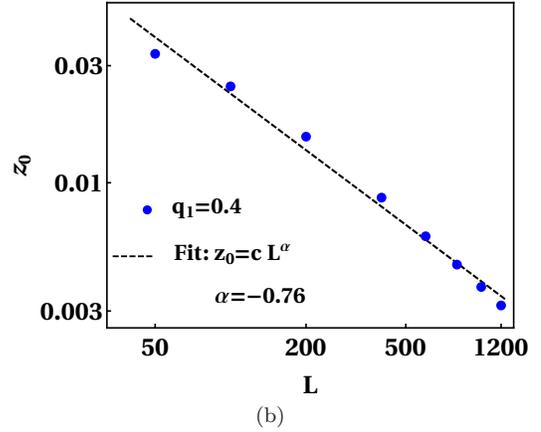}}\\
\caption{The figure (a) illustrates the convergence of the fugacity behavior with increasing system size $L$, plotted against the parameter $q_1$. The figure (b) how the approach of $z_0$ towards this convergence with varying $L$, at the transition point $q_1=p_1$. The exact numerical data obtained from the solution of $z_0$ is fitted with the form $z_0=c L^\alpha$, resulting in $\alpha\approx-0.76$. The parameters used are $p_1=0.4, p_2=1.0, q_2=0.5, \epsilon=0.1, w_{12}=1.0, w_{21}=0.1, \rho_+=0.25, \rho_0=0.4$.}.
\label{fig:zvsl}
\end{figure}
\begin{figure}[t]
\centering
\subfigure[]{\includegraphics[scale=0.47]{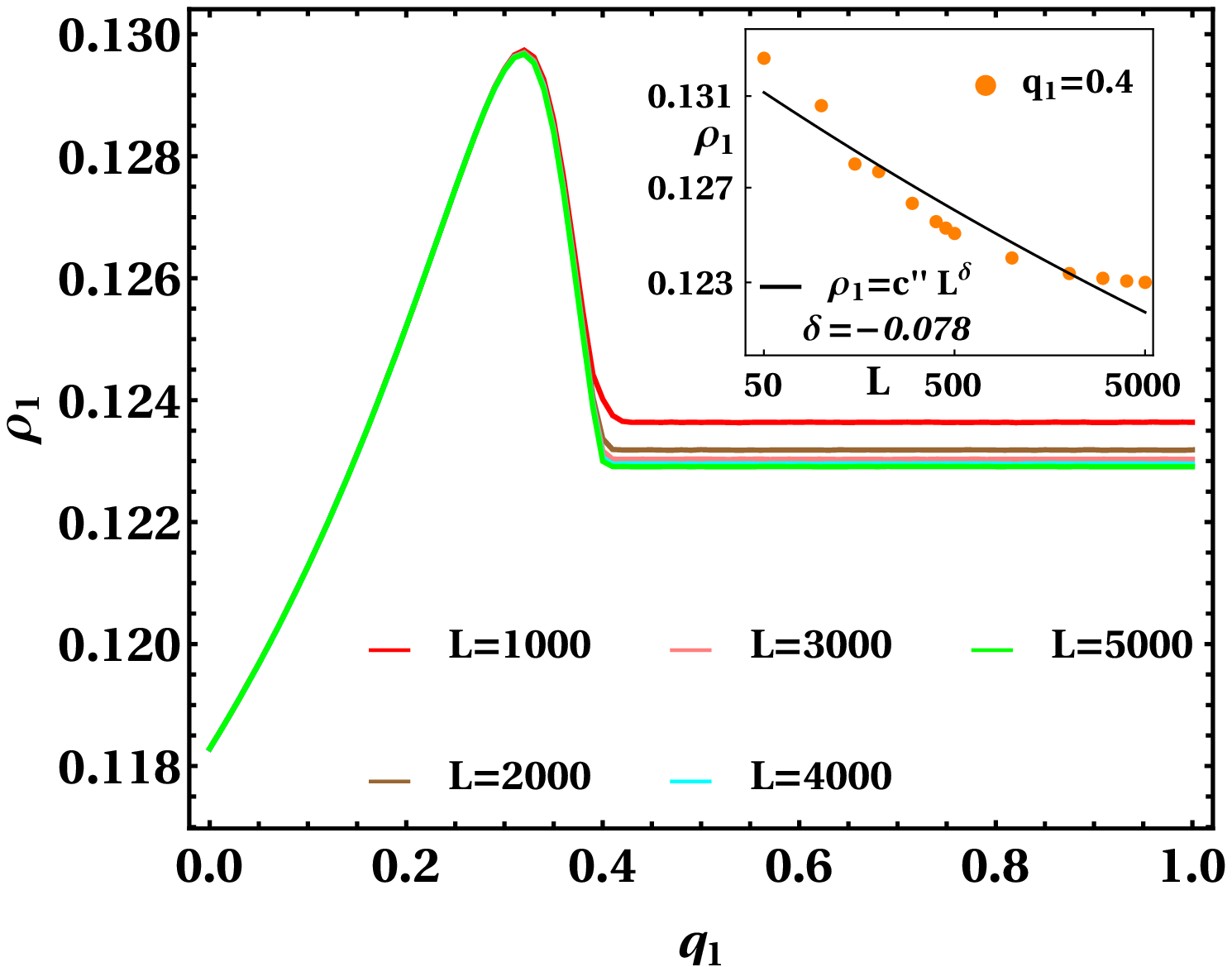}}\\
\subfigure[]{\includegraphics[scale=0.47]{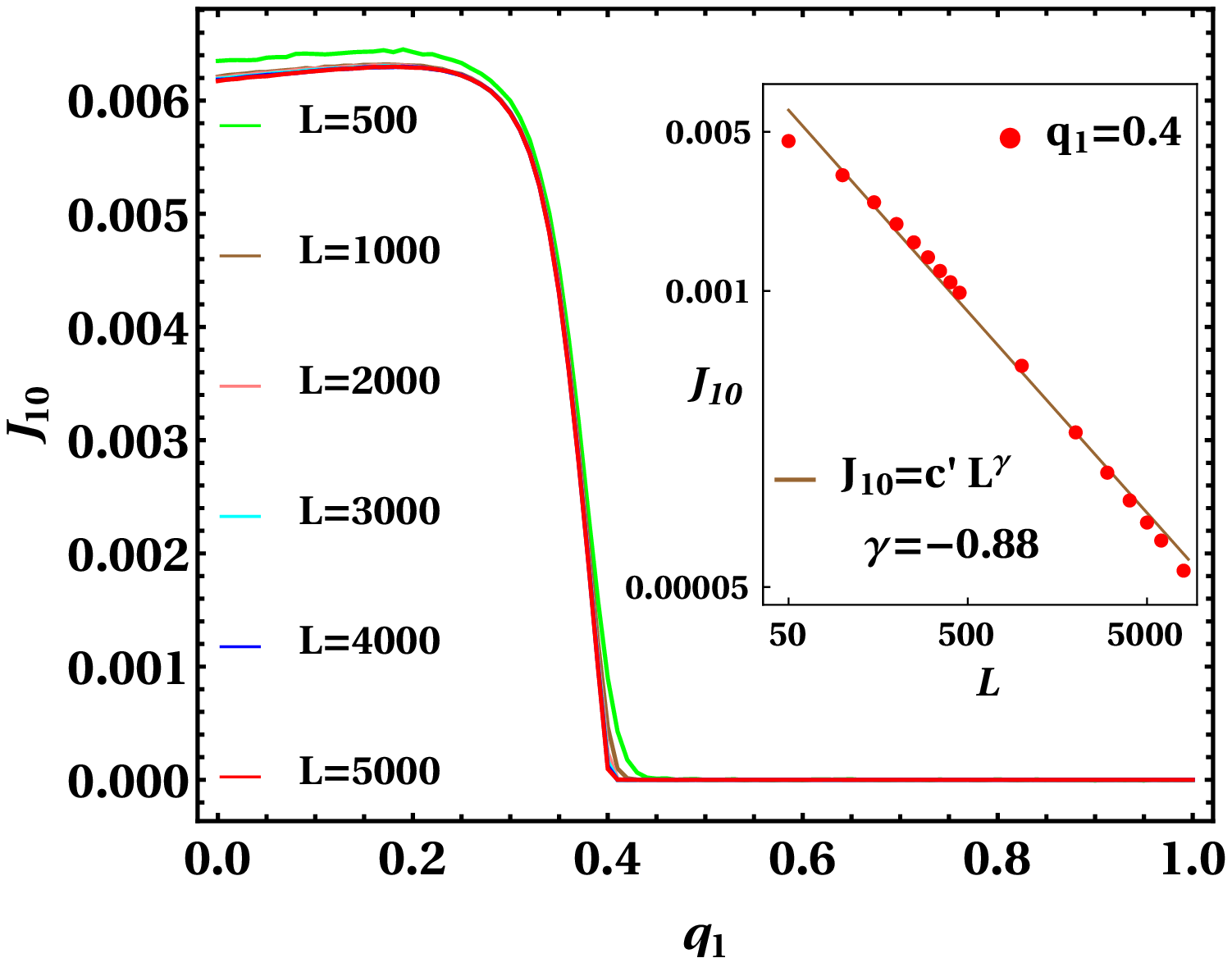}}\\
\caption{The figures (a) and (b) show convergence of $\rho_1$ and $J_{10}$ respectively, with increasing system size $L$. The insets (in log scale) of (a) and (b) show the variations of $\rho_1$ and $J_{10}$ respectively with $L$ at the transition point $q_1=p_1$. They follow the relations $\rho_1=\bar{\rho}c''L^\delta$ and $J_{10}=c'L^\gamma$ with exponent values $\delta\approx-0.078$ and $\gamma\approx-0.88$ obtained from fitting. The parameters used are $p_1=0.4, p_2=1.0, q_2=0.5, \epsilon=0.1, w_{12}=1.0, w_{21}=0.1, \rho_+=0.25, \rho_0=0.4$.}.
\label{fig:l}
\end{figure}
Note that, the density $\rho_1$ should be an intensive quantity independent of $L$ and the steady state current should also be a constant independent of $L$. In fact, like the fugacities e.g. temperature, chemical potential which are independent of system size, the fugacity $z_0$ should also be a constant independent of system size. Thus, $z_0$, $\rho_1$ and $J_{10}$ here must show convergence with increasing $L$. Indeed, the fugacity obtained from the exact numerical data by solving Eq.~(\ref{eq:srhozexp}), show converging behavior with increasing system size in Fig.~\ref{fig:zvsl}(a). At the transition point $q_1=p_1$, we investigate the approach of $z_0$ towards convergence as a function of $L$, by fitting the data with the functional form $z_0=c L^\alpha$ in Fig.~\ref{fig:zvsl}(b). Also, both Figs.~\ref{fig:l}(a) and (b) exhibit convergence of the corresponding behaviors of $\rho_1$ and $J_{10}$ with increasing system size. In Fig.~\ref{fig:l}(b), the current shows rapid convergence in both the free flowing phase and clustering phase for $L=10^3$ and more. Notably, in Fig.~\ref{fig:l}(a), the density shows faster convergence with increasing $L$ in the free flowing phase, in comparison to that of the clustering phase. We should mention that the convergence value of $\rho_1$ in the clustering phase in Fig.~\ref{fig:l}(a), agrees with the simplified formula obtained in Eq.~(\ref{eq:rho-constant}). In the insets of Figs.~\ref{fig:l}(a) and (b), we examine the approach of $\rho_1$ and $J_{10}$ respectively towards convergence, with varying system size $L$. We fit the corresponding data with the relations of the form $\rho_1=c'' L^\delta$ and $J_{10}=c'L^\gamma$. Keeping in mind the relation [Eq.~(\ref{eq:j})] between current, density and fugacity of the form $J_{10}=z_0 \rho_1$, the exponent $\delta$ can be determined from the relation $\delta=(\gamma-\alpha)$. Our findings in Figs.~\ref{fig:zvsl} and \ref{fig:l}(b) predict $\delta$ to be $(\gamma-\alpha)=-0.12$. This is close to but not exactly equal to the $\delta$ value ($\approx-0.078$) obtained in the inset of Fig.~\ref{fig:l}(a), because of the finite system sizes used in the discussion.
\section{Arndt-Heinzel-Rittenberg model of counter-flow: comparisons}
\label{app:ahr}
In this section, we compare the microscopic dynamics of our model with the dynamics of the  
Arndt-Heinzel-Rittenberg (AHR) model that is known to exhibit three different phases \cite{Arndt_1998,sArndt_1998_1,sArndt_1999}. The AHR model considers positive (say, species $1$) and negative (say, species $2$) particles along with vacancies ($0$) on a one dimensional periodic lattice and the follow the dynamical rules given below
\begin{eqnarray}
 10 \,\,\stackrel{\lambda}{\longrightarrow}\,\,  01,  \hspace*{0.8 cm} 
 02 \,\,\stackrel{\lambda}{\longrightarrow}\,\,  20,  \hspace*{0.8 cm} 
12\,\, \xrightleftharpoons[1]{q}\,\, 21 .
 \label{eq:sahr}
\end{eqnarray}
A straightforward comparison of the AHR model in Eq.~(\ref{eq:sahr}) with our model in Eq.~(\ref{eq:dynamics}) reveal the following factors: (i) our model allows a {\it non-conserving flip dynamics} that activates transformations between species $1$ and $2$, which is absent in the AHR model for which each microscopic dynamics maintains particle number conservation of every species. Also, as a consequence, our model requires minimum four species in total (species $1$, species $2$, impurity and vacancy) to operate both drift and flip dynamics, while the AHR model deals with three species in total (species $1$, species $2$, and vacancy). (ii) AHR model has an {\it exchange dynamics} that allows the two species to exchange their positions. Such exchange dynamics is absent in our case. (iii) Our model is {\it non-ergodic} in contrast to the {\it ergodic} nature of the AHR model. 

The exchange rate $q$ is considered as the tuning parameter for the AHR model and the density $\rho$ of the two species are taken to be equal \cite{Arndt_1998,sArndt_1999}. Three different phases, namely {\it pure phase}, {\it mixed phase} and {\it disordered phase} are observed, as $q$ is varied \cite{Arndt_1998,sArndt_1999}. When $q<1$, the species $1$ is more probable to reside at left of species $2$ which again likes to be at left of vacancies. This leads to a complete species segregation with three types of blocks each purely consisting of one species (either $1$ or $2$ or $0$), thereby referred as pure phase. For $1<q<q_c$ (where $q_c$ depends on $\lambda$ and $\rho$), a condensate is formed that has both species $1$ and species $2$ mixed up, accompanied by a fluid consisting of vacancies and some particles of the two species. This phase is known as the mixed phase. There is no species segregation or condensate formation for $q>q_c$, which is the disordered phase. To observe these three phases, two point functions like drift current and correlations between different species, have been used \cite{Arndt_1998,sArndt_1999}.

Interestingly, in our model, due to the presence of non-conserving flip dynamics, we have even simpler one point function like average species densities  among observables of interest, and indeed the average species densities clearly show the existence of two different phase, the {\it free flowing phase} and the {\it clustering phase}. The free flowing phase in our model is similar to the disordered phase of AHR model. On the other hand, for the specific choice of initial configuration considered here and due to the flip dynamics, species segregation in {\it pure} form is not possible in the clustering phase. Rather, we have two macroscopic clusters, one consisting of only vacancies and the other consisting of all kinds of particles (species $1$ and species $2$ and impurities). Although the mixing up of different species and impurities inside the particle cluster has resemblance to the mixed phase condensate of AHR model, we do not have a background fluid in our case. Rather, for any $q_1>p_1$, we have only two clusters in the clustering phase, a vacancy cluster and a particle cluster, with the particle cluster drifting along right with considerably small velocity that depends on the density of the non-flipping species $1$ particles in the system. Another noteworthy point in our analysis is the {\it rearrangement parameter} whose variation captures the effect of non-ergodicity on the clustering phenomenon, there is no such counterpart in the ergodic AHR model. 

Note that exact analysis in Refs.~\cite{Rajewsky_2000} and \cite{Sasamoto_2001} later revealed that there is actually no phase transition between mixed and disordered phase in the AHR model, in the thermodynamic limit within the grand canonical ensemble framework. This conclusion is associated with the existence of extremely long but still finite correlation lengths in the infinite system.

There is an alternative approach to compare the dynamics of the two models, although the key points of the comparative analysis between them remain the same. Our tuning parameter has been the counter-flow parameter $q_1$, which is a part of the drift dynamics. To treat the tuning parameter $q$ of the AHR model on an equivalent footing, one can relabel species $1$, species $2$ and vacancy in AHR model as vacancy, species $1$ and species $+$ respectively. Thus one arrives at an alternative version of the AHR model given below
\begin{eqnarray}
 0+ \,\,\stackrel{\lambda}{\longrightarrow}\,\,  +0,  \hspace*{0.8 cm} 
 +1 \,\,\stackrel{\lambda}{\longrightarrow}\,\,  1+,  \hspace*{0.8 cm} 
10\,\, \xrightleftharpoons[q]{1}\,\, 01 .
 \label{eq:sahralt}
\end{eqnarray}
Here we see, in comparison to our model Eq.~(\ref{eq:dynamics}), the species $2$ is absent and we do not term $+$ as impurity because there is no flip dynamics at all. So $+$ drifts to left only, species $1$ drifts to right or left with rates $q$ and $1$ respectively, with an additional exchange of positions between $+$ and $1$. Clearly $q>1$ here corresponds to natural flow situation and $q<1$ refers to the counter-flow situation.

The exact steady state of AHR model in Eq.~(\ref{eq:sahr}) has been obtained in matrix product from, which has  a two dimensional representation in the limit $q\rightarrow\infty$ and infinite dimensional representations in general \cite{sArndt_1999}, which have different structures in comparison to the infinite dimensional matrices in our case Eq.~(\ref{eq:matrices}).

\end{document}